\documentclass[fleqn,usenatbib]{mnras}

\usepackage[T1]{fontenc}
\usepackage{ae,aecompl}

\usepackage{graphicx}	
\usepackage{amssymb}	
\usepackage{savesym}
\usepackage{amsmath}    
\savesymbol{iint}
\usepackage{txfonts}
\restoresymbol{TXF}{iint}
\allowdisplaybreaks

\title[HFQPOs in eccentric discs]
{HFQPOs and discoseismic mode excitation in eccentric, relativistic discs. I. Hydrodynamic simulations}

\author[J. Dewberry et al.]{
Janosz W. Dewberry,$^{1,2,3}$\thanks{E-mail: janosz.dewberry@cantab.net}
Henrik N. Latter,$^{3}$
Gordon I. Ogilvie,$^{3}$
Sebastien Fromang$^{4,5}$
\\
$^{1}$Tsung-Dao Lee Institute, No. 800 Dongchuan Road, Minhang District, Shanghai 200240, China\\
$^{2}$Department of Astronomy, Center for Astrophysics and Planetary Science, Cornell University, Ithaca, NY 14853, USA\\
$^{3}$DAMTP, University of Cambridge, CMS, Wilberforce Road, Cambridge, CB3 0WA, UK\\
$^{4}$Laboratoire AIM, CEA/DSM-CNRS-Universit\'e Paris 7, Irfu/Departement d'Astrophysique, CEA-Saclay, F-91191 Gif-sur-Yvette, France\\
$^{5}$Laboratoire des Sciences du Climat et de l'Environnement, LSCE/IPSL, CEA-CNRS-UVSQ, Universit\'e Paris-Saclay, F-91190 Gif-sur-Yvette, France
}

\date{Accepted XXX. Received YYY; in original form ZZZ}

\pubyear{2020}

\begin{document}
\label{firstpage}
\pagerange{\pageref{firstpage}--\pageref{lastpage}}
\maketitle

\begin{abstract}
High-frequency quasi-periodic oscillations (HFQPOs) observed in the emission of black-hole X-ray binary systems promise insight into strongly curved spacetime. `Discoseismic' oscillations with frequencies set by the intrinsic properties of the central black hole, in particular `trapped inertial waves' (r-modes), offer an attractive explanation for HFQPOs. To produce an observable signature, however, such oscillations must be excited to sufficiently large amplitudes. Turbulence driven by the magnetorotational instability (MRI) fails to provide the necessary amplification, but r-modes may still be excited via interaction with accretion disc warps or eccentricities. We present 3D global hydrodynamic simulations of relativistic accretion discs, which demonstrate for the first time the excitation of trapped inertial waves by an imposed eccentricity in the flow. While the r-modes' saturated state depends on the vertical boundary conditions used in our unstratified, cylindrical framework, their excitation is unambiguous in all runs with eccentricity $\gtrsim 0.005$ near the ISCO. These simulations provide a proof of concept, demonstrating the robustness of trapped inertial wave excitation in a non-magnetized context. In a companion paper, we explore the competition between this excitation, and damping by magnetohydrodynamic turbulence.
\end{abstract}




\section{Introduction}
X-ray binaries comprising an accreting black hole and a donor star (hereafter BHBs) progress through distinct emission states during outburst, on time-scales of weeks to months. These states are characterized, in part, by the relative dominance of a `thermal' blackbody (i.e., disc-like) emission component with a characteristic temperature of $\sim 1$keV, and a harder `power-law' component at higher energies. During `steep-power law' (SPL) or `very high' emission states, in which both of these emission components are strong, BHBs occasionally exhibit `high-frequency quasi-periodic oscillations' (HFQPOs). While subtle, infrequent, and poorly understood, HFQPOs excite considerable theoretical interest: their frequencies of $\sim50-450$Hz (i) are comparable with the orbital and epicyclic frequencies expected in the very inner regions of a relativistic accretion disc, (ii) are relatively insensitive to variations in luminosity, and (iii)  scale inversely with estimates of black hole mass. These characteristics suggest that a robust model for HFQPOs may provide measurements of, in particular, black hole spin angular momentum \citep{Rem06,Done07,Bel12,bel16,Mot16}. 

Nearly every model for HFQPOs appeals to the effects of general relativity on orbital motion, specifically the horizontal and vertical epicyclic frequencies $\kappa$ and $\Omega_z$, both of which deviate from the orbital frequency $\Omega$ near the black hole. The `relativistic precession model' \citep[RPM; ][]{Stel98,Stel99} takes perhaps the simplest approach, associating quasi-periodic oscillations with the orbital frequency and both the nodal and periastron precession frequencies $\Omega-\Omega_{z}$ and $\Omega-\kappa$ \citep[e.g.,][]{Mot14}. The appearance of multiple HFQPOs with frequencies in near-integer ratios in some BHBs (in particular GRO J1655-40) has led to the development of models that rely on resonances between fluid `blobs' orbiting at radii where $\kappa$ and $\Omega_z$ achieve the same commensurability \citep{Kluz01,Abr01}, or the global acoustic oscillations of narrow, pressure-supported accretion tori \citep{Rez03,Blaes06,Horak08,Frag16}. 

The model considered in this paper posits that HFQPOs may be identified with global waves excited in relativistic thin, \emph{radially extended} discs. As first recognized by \cite{Oka87}, general relativistic effects close to a black hole modify the radial profile of $\kappa$ in such a way that inertial oscillations (r-modes) can be trapped in a wave cavity. Trapped r-modes possess frequencies close to the maximum epicyclic frequency, and semi-analytic theory predicts they can be amplified by a non-linear coupling with any eccentricity and/or warp in the accretion flow \citep{Kat04,Kat08,fer08}. Eccentricities arise naturally from the tidal influence of the companion star, while warps result from the misalignment of the disc's orbit and the black-hole spin angular momentum.  Importantly, the large accretion rates associated with the very high state (in which HFQPOs are observed) can aid the propagation of such distortions to the inner disc \citep{fer09}. 

In this paper, we demonstrate the excitation of trapped r-modes in fully non-linear, three-dimensional numerical simulations, and take some steps toward describing their ensuing saturated state. We focus here on the unmagnetized problem, and on eccentric (rather than warped) discs. This hydrodynamic work provides a point of comparison for magnetohydrodynamic (MHD) simulations of turbulent discs considered in a companion paper.

Using the code RAMSES, we run simulations that utilize a pseudo-Newtonian Paczynski-Wiita potential, and which omit vertical gravity for simplicity. We permit material to flow through an inner boundary placed \emph{within} the innermost stable circular orbit (ISCO), and so our simulations contain a physical disc edge and a transonic plunging region. This plunging region is more physically motivated for a black hole accretion disc, and \citep[as noted by ][]{mir15}, offers less reflection to spiral density waves excited via the corotation instability considered by \cite{lai09}, \cite{FL11} and \cite{FL13}. Meanwhile, we impose a non-axisymmetric pressure gradient at the outer boundary that forces the inward propagation of eccentricity and the formation of a twisted eccentric disc composed of misaligned elliptical orbits. This structure approximates an eccentric wave forced in the outer disc by the companion \citep{fer09}.

We find that simulations exhibiting even very low values of eccentricity ($\gtrsim0.005$) witness the excitation of trapped inertial waves. In the presence of rigid vertical boundaries they take the form of global standing modes, while periodic vertical boundaries lead to a saturated state involving steady vertical `elevator' flows. The r-modes produce peaks in the power spectral density (PSD) in a narrow range encompassing the frequency (approximately the epicyclic) and radii predicted by the analytic theory. Through a non-linear coupling at larger amplitudes, the trapped inertial waves additionally excite secondary inertial-acoustic modes at nearly double their frequency. Meanwhile, the waves rearrange angular momentum locally and hence dynamically reshape the trapping region in which they are confined. As a consequence, their frequencies vary in time, possibly contributing to the low quality factors of observed HFQPOs.

The paper takes the following structure. In Section \ref{sec:bgrnd} we provide theoretical background, while details of the numerical framework are discussed in Section \ref{sec:num}. We present our simulations exhibiting r-mode excitation in Section \ref{sec:3D}, before concluding in Section \ref{sec:conc}. Finally, test simulations and a resolution study are discussed in the appendices.

\section{Theoretical background}\label{sec:bgrnd}

In this section, we provide an introduction to discoseismology, describe the non-linear excitation mechanism responsible for r-mode amplification, and discuss problems and complications associated with the discoseismic model for HFQPOs. Those already familiar with this theoretical background may skip to Sections \ref{sec:num} and \ref{sec:3D}.

\subsection{Discoseismic oscillations}

Consider isothermal, hydrodynamic perturbations to a thin, purely rotating,  barotropic fluid disc with angular velocity $\Omega=\Omega(r)$. Such linear disturbances are governed by the local dispersion relation \citep{Oka87}
\begin{equation}\label{eq:hdisp}
    k_r^2=\frac{(\hat{\omega}^2-\kappa^2)(\hat{\omega}^2-n\Omega_z^2)}{\hat{\omega}^2c_s^2}.
\end{equation}
Here $k_r$ is the radial wavenumber of the perturbation, $c_s$ is the isothermal sound speed, and $\hat{\omega}=\omega-m \Omega$ for $\omega$ the (complex) frequency and $m$ the azimuthal wavenumber. Meanwhile, $\kappa$ and $\Omega_z$ are once again the horizontal and vertical epicyclic frequencies. Finally, $n$ is a separation constant associated with the decomposition of the perturbations' vertical structure in terms of Hermite polynomials, and describes the number of nodes in the vertical direction. In a cylindrical model excluding vertical gravity and density stratification, the perturbations' vertical structure may be described by a vertical wavenumber $k_z$, and $n\Omega_z^2$ is replaced by $k_z^2c_s^2$. 

Equation \eqref{eq:hdisp} is derived from Newtonian hydrodynamics, but permits the approximate inclusion of relativistic effects on wave propagation through the characteristic frequencies $\kappa$ and $\Omega_z$. A common practice is to calculate these frequencies from a `pseudo-Newtonian' Paczynski-Wiita potential given by Equation \eqref{PWpot}. In a centrifugally supported disc, the radial profile for the horizontal epicyclic frequency so calculated shares two important properties with its counterpart in full general relativity \citep{Oka87}. First of all, both radial profiles for $\kappa^2$ pass through zero at a radius defining the innermost stable circular orbit; within this radius, fluid elements or particles are unstable to horizontal displacements. Second, the combination of $\kappa$'s asymptotic decay at large $r$ ($\kappa\propto r^{-3/2}$) with the existence of an ISCO implies that $\kappa$ possesses a maximum. This non-monotonic behavior, illustrated by the black dashed line in Fig. \ref{fig:hvprop}, is the source of a relativistic disc's capacity to support discrete modes of oscillation irrespective of radial boundary conditions. 

\subsubsection{Inertial-acoustic oscillations (f-modes)}
\begin{figure}
    \centering
    \includegraphics[width=\columnwidth]{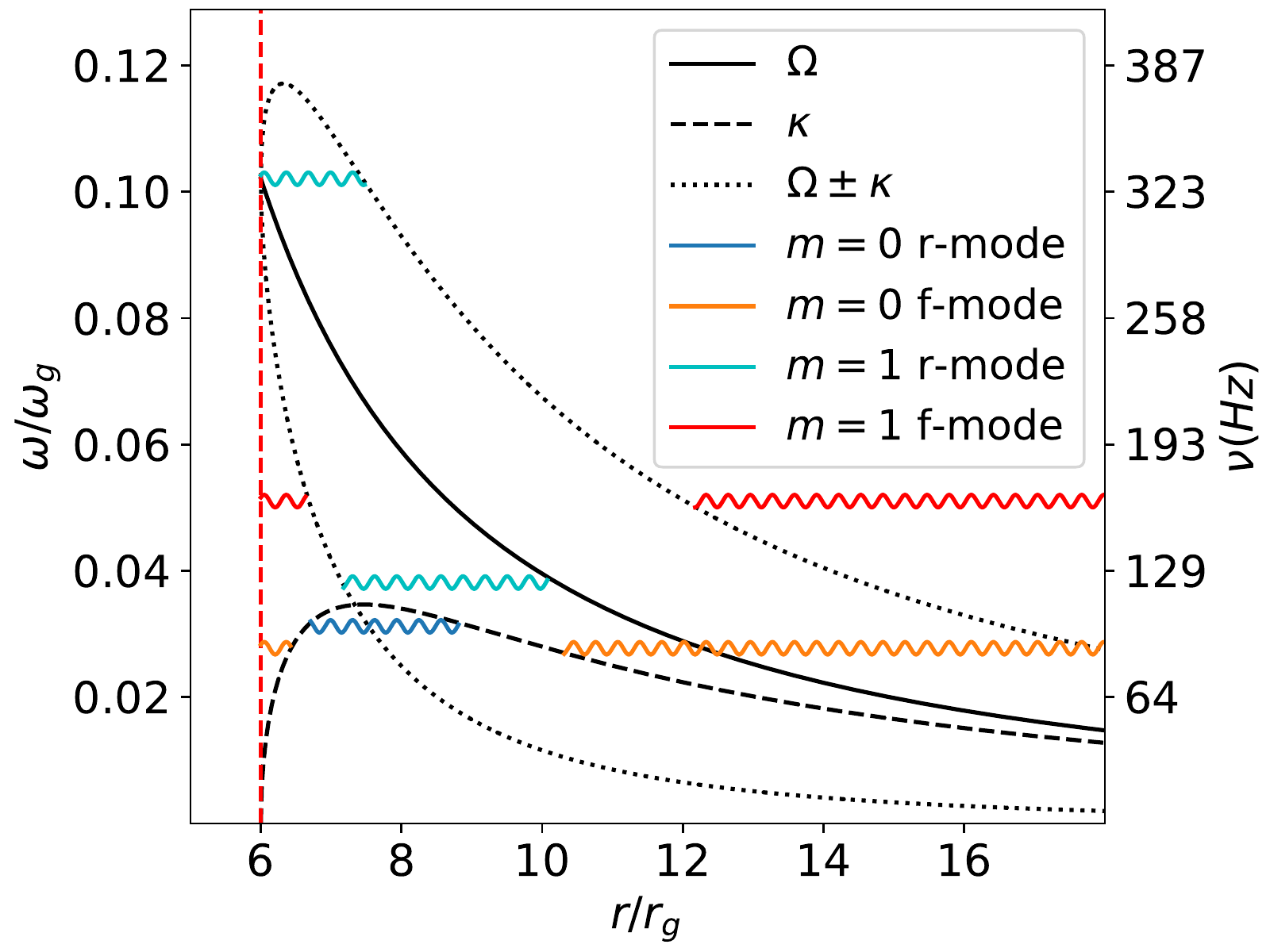}
    \caption{Schematic wave propagation diagram illustrating the radii of localization for both axisymmetric ($m=0$) and non-axisymmetric ($m\not=0$) inertial and inertial-acoustic waves (r-modes and f-modes, resp.). The sinusoidal lines indicate the regions within which the modes of a given character and azimuthal structure are predicted to be oscillatory by dispersion relation \eqref{eq:hdisp}. Their intersections with the characteristic frequencies $\Omega,$ $\kappa,$ and $\Omega\pm \kappa$ define resonant radii beyond which the oscillations become evanescent. Here $r_g$ and $\omega_g$ are the gravitational radius and frequency (see Section \ref{sec:phys}), and frequencies $\nu$ given in Hz are calculated for a $10M_\odot$ black hole using a pseudo-Newtonian, Pacynski-Wiita potential.}
    \label{fig:hvprop}
\end{figure}

Equation \eqref{eq:hdisp} predicts several families of oscillatory, non-evanescent modes with $k_r^2>0$. The simplest, vertically homogeneous ($n=0$) `inertial-acoustic' modes (often called density waves or even p-modes, but hereafter in this work called f-modes) can propagate where $\hat{\omega}^2>\kappa^2.$ \cite{kat80} posited that in a relativistic disc, axisymmetric inertial-acoustic waves with $m=0$ and a frequency $\omega\lesssim\max[\kappa]$ could be trapped between the ISCO and the peak in epicyclic frequency, if reflected at the disc edge. This trapping is illustrated by the orange wiggly lines in Fig. \ref{fig:hvprop}. More generally, \emph{non-axisymmetric} inertial-acoustic modes with $m\not=0$ can propagate where 
$\omega>m \Omega+\kappa$ or $\omega<m \Omega-\kappa$ (see the red wiggly lines in Fig. \ref{fig:hvprop}). 

Trapped f-modes feature in several HFQPO models, because it is possible for them to be amplified by various instabilities. They can become viscously overstable when, for instance, the disc possesses a transonic radial inflow through the ISCO \citep{kat78,chan09,mir15}. Alternatively, in discs with a reflecting inner boundary, non-axisymmetric f-modes may be excited via a transmission of wave energy across their corotation radii (where $\omega/m$ matches $\Omega$). This excitation mechanism appeals to the same principles underlying the Papaloizou-Pringle instability \citep{Papri}; non-axisymmetric f-modes propagating within corotation can be seen as carrying negative wave energy, and grow in amplitude if they transmit positive wave energy to the outer disc. In relativistic discs, a positive gradient in fluid `vortensity' $\kappa^2/(2\Sigma\Omega)$ at the corotation radius facilitates such transmission (here $\Sigma$ is the surface density) \citep{lai09,FL11,FL13}. However, the energy exchange excites large-scale normal modes only if the inner boundary reflects the oscillations \citep{mir15}. Such reflection might be caused by, e.g., a magnetosphere or a neutron star's surface, but is less likely in an accretion disc around a black hole delimited by a plunging region.

\subsubsection{Trapped inertial oscillations (r-modes)}
In contrast, vertically structured oscillations can exist independently of the inner boundary. For non-zero $n$ (or $k_z$) dispersion relation \eqref{eq:hdisp} describes higher frequency acoustic p-modes, which propagate where  $\hat{\omega}^2>\max[\kappa^2,n\Omega_z^2]=n\Omega_z^2,$ and lower frequency inertial r-modes, localized to radii where $\hat{\omega}^2<\min[\kappa^2,n\Omega_z^2]=\kappa^2$. Trapped inertial waves are frequently referred to as g-modes for the mathematical similarity of their trapping to that of internal gravity waves in stars, but we use the name r-modes to emphasize that they are restored by the Coriolis force, rather than buoyancy. Axisymmetric ($m=0$) r-modes are of the greatest interest for discoseismic theories of HFQPOs, as the maximum in $\kappa^2$ serves as a `self-trapping region,' providing inner and outer turning points at the radii where $\omega^2=\kappa^2$ (as illustrated by the dark blue wiggly line plotted in Fig. \ref{fig:hvprop}). In addition to eliminating the need for reflection at the inner boundary, confinement to a narrow annulus mitigates any damping by radial inflow. Importantly, the frequencies of these waves track the maximum of the epicyclic frequency and therefore, when matched to observations, might be used as diagnostics of black hole mass and spin. 

The cyan lines in Fig. \ref{fig:hvprop} also indicate the regions of confinement for non-axisymmetric r-modes with azimuthal wavenumber $m=1.$ Such oscillations may be similarly confined away from radial boundaries, but have been disfavoured as an explanation for HFQPOs because they are strongly damped at corotation \citep{Li02,lat09}. Non-axisymmetric r-modes with frequencies large enough that their corotation radius lies outside of their trapping region (as illustrated by the higher frequency cyan line in Fig. \ref{fig:hvprop}) might avoid this damping, but would then have frequencies too high to explain HFQPOs in systems containing black holes with any non-negligible spin angular momentum \citep{Wag12}. On the other hand, as described in the next section, lower frequency non-axisymmetric inertial waves lying within their corotation radius do play an ancillary role in the excitation mechanism for $m=0$ r-modes considered in this paper \citep{Kat04,Kat08,fer08,okt10}.

There have been a handful of attempts to match the frequencies produced by the linear wave theory to those associated with observed HFQPOs (\citealp{Wag01,Wag12}, see also \citealp{Kat08b}). Though these studies are broadly consistent, several issues complicate a direct comparison of frequencies. Trapped inertial mode frequencies can be altered by poloidal magnetic fields or strong pressure gradients \citep{FL09,dew18}. As we discuss in Section \ref{sec:mods}, trapped modes of sufficient amplitude appear to alter the background angular momentum distribution, in turn reshaping their own trapping regions and, as a consequence, their frequencies. The non-linearity inherent in such self-interaction may also bear on the problem presented by some sources that exhibit \emph{multiple} HFQPOs with distinct frequencies. One of the motivations of the present numerical study is to gain a better understanding of this non-linearity.

Finally, we note that fully general relativistic expressions for $\Omega$, $\kappa$ and $\Omega_z$ have qualitatively similar radial profiles but different frequencies than those shown in Fig. \ref{fig:hvprop}. Importantly, the maximum in $\kappa$ (and hence r-mode frequencies) can be significantly larger for a non-zero black hole spin angular momentum. Consequently the frequencies we generate in this and our companion paper, though possessing the correct qualitative behaviour, are not quantitatively correct and should not be compared directly to observations.

\subsection{Excitation of r-modes by warps and eccentricities}\label{sec:bgEx}

While acoustic modes can be excited by convection in stars, MHD turbulence driven by the magnetorotational instability (MRI) fails to excite trapped inertial modes in relativistic accretion discs \citep[][see Section \ref{sec:mag}]{ReM09}. The absence of turbulent excitation motivates consideration of alternative r-mode amplification processes; we now review the most promising mechanism, which involves large-scale disc deformations.

\cite{Kat04,Kat08}, \cite{fer08} and \cite{okt10} discovered and elucidated a process by which a disc distortion, such as a warp or eccentric structure, facilitates energy exchanges between r-modes and the disc, so that the r-modes grow exponentially. The mechanism is similar to the well-known parametric instability in stars and discs, in which a `parent' mode transfers energy to pairs of `daughter' modes. Locally, disc deformations can certainly trigger parametric instability, with the distortion serving as the parent mode from which two daughter waves (travelling inertial waves) extract energy \citep{goo93,gam00,pap05a,pap05b,ogi13,bar14}. The instability mechanism for trapped r-modes in relativistic discs is more complicated, as discussed by \cite{fer08}: in this context the ultimate source of energy is the background differential rotation, not the deformation itself, which instead permits the three-wave coupling that can then draw out rotational energy. 

The three waves that take part in the instability mechanism consist of (i) the non-axisymmetric eccentricity (or warp), characterized by wavenumbers $m_E=1,n_E=0$ (or $m_W=1$, $n_W=1)$, (ii) the fundamental, axisymmetric trapped r-mode with $m_R=0,n_R=1,$ and (iii) an intermediate \textit{non-axisymmetric} wave with $m_I=1,n_I=1$ (or $n_I=1\pm 1$), propagating within its corotation radius. The three coupled modes must propagate over the same disc radii and satisfy certain coupling rules. Since the disc warp or eccentricity has nearly zero frequency, and the trapped r-mode has a frequency just less than $\max[\kappa]$, the intermediate wave has almost the same frequency. The wiggly dark blue and (lower frequency) cyan curves in Fig. \ref{fig:hvprop} represent (approximately) two inertial oscillations that possess matching frequencies, and thus can couple via the disc deformation. Now, since the intermediate $m=1$ inertial wave propagates within corotation, it carries negative wave energy, in that its presence reduces the total energy of the disc. As a consequence, the axisymmetric r-mode (which carries positive wave energy) is amplified, and any damping of the intermediate wave (such as it might encounter at corotation or via turbulent dissipation) will continuously draw orbital energy from the background disc and channel it into the trapped axisymmetric r-mode \citep{Kat04,Kat08,fer08}.

While the calculations of \cite{fer08} demonstrated robust growth rates for even small values of eccentricity, the three-wave coupling considered by the authors assumes that the axisymmetric inertial mode and intermediate inertial wave are sufficiently small in amplitude that they interact only via the disc distortion.  In other words, the authors employed a theory that is linear in the growing mode. Moreover, they considered disc deformations small enough in amplitude that they could be represented as linear normal modes, and taken as time-independent. The instability may take on a different character for stronger warps and eccentricities, and it remains to be validated in fully non-linear simulations, or considered in conjunction with MHD turbulence.

\subsection{Inward eccentricity propagation}
Discoseismology's appeal to r-modes as the origin of HFQPOs rests on the excitation mechanism described in Section \ref{sec:bgEx} --- and hence on a sufficiently strong disc distortion in the innermost radii. It is uncontroversial that many BHBs exhibit a misalignment between the angular momentum of the accretion flow and that of the compact object, which will produce a warp in the inner disc \citep{bar75}. But it is perhaps less appreciated that an eccentricity generated in the outer disc can propagate into the inner disc relatively unimpeded, under certain circumstances. For this reason, and because the simulations in this paper deal with the more manageable problem of eccentric discs, we now review the theory of eccentricity propagation.

Several pieces of evidence point to significant eccentricities in the outer radii of many BHBs. Discs in close binary systems with $M_2/M_1\lesssim0.3$ (typical in BHBs) can be subject to an eccentric instability associated with a 3:1 resonance between the disc and the binary orbit \citep{lub91a,lub91b,lub10}. Observationally, this manifests most clearly as long-period modulations of emissivity (`superhumps') in SU UMa systems, which are associated with the precession of eccentric discs that are tidally distorted by the companion \citep{whi88,pat05}. Large eccentricities of $\sim0.1-0.3$ have been measured in the outer parts of SU UMa discs \citep[e.g.,][]{hes92,pat00,pat02,rol01}, in rough agreement with numerical simulations \citep[see ][ and references therein]{kle08}. Superhumps are also observed in low-mass X-ray binaries \citep{odo96,nei07,zur08,kos18}, indicating that a similar excitation of eccentricity takes place in the outer regions of BHB accretion discs. 

\begin{figure}
    \centering
    \includegraphics[width=\columnwidth]{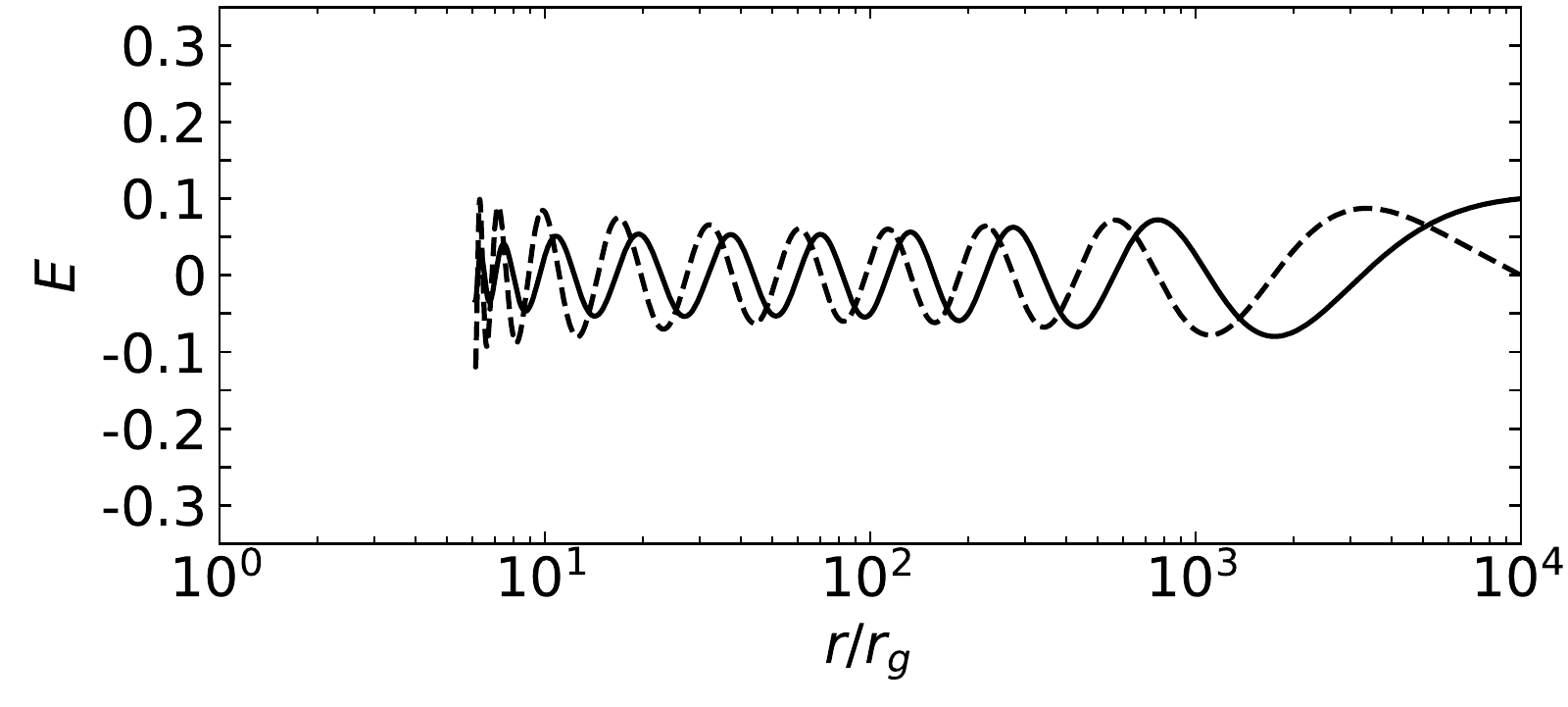}
    \caption{The real (solid) and imaginary (dashed) parts of the complex eccentricity for an inward travelling eccentric wave as calculated by \citet{fer09}. The accretion rate is $\dot{M}/\dot{M}_\mathrm{Edd}=0.5$. The eccentricity at the outer radius is 0.1, the $\alpha$ parameter is 0.1, and the turbulent damping is parameterised with a bulk viscosity with $\alpha$ also of 0.1.}
    \label{fig:eprop}
\end{figure}

An important question is then how deeply into the disc such outer eccentricities can propagate, especially in the presence of turbulent damping. \cite{fer09} offered the first attempt at an answer, demonstrating that eccentricities manifest as very long-wavelength disturbances at large radii, and only steepen to wavelengths on par with the disc scale height very close to the ISCO. These long-wavelength disturbances should experience minimal damping by small-scale turbulence throughout the majority of the disc, and can reach the ISCO relatively undiminished if the mass accretion rate is large, as is the case in the SPL state.
When the accretion rate is lower, however, there can be significant attenuation of the distortion.
Fig. \ref{fig:eprop} presents an illustrative calculation of the complex eccentricity $E=e\exp[\text{i}\varpi]$ as a function of radius, where $e$ is eccentricity and $\varpi$ is the longitude of pericentre. The plot shows that the inward-propagating eccentric wave maintains its amplitude from $r=10^4 r_g$ all the way to the ISCO.

The calculations of \cite{fer09} made use of a linear secular theory, but recently \citet{ogi19} and \citet{lyn19} developed a fully non-linear Hamiltonian formalism that revised, in part, these earlier results. They found that non-linear effects can lead to a decrease of disc eccentricity (relative to linear predictions) near the ISCO, on the one hand, and the likely formation of shocks in the same inner regions, on the other, though the associated dissipation was not included in the theory. It is probable that these non-linear effects control the amplitude of the eccentric wave as it approaches the ISCO more than small-scale turbulent damping. In this paper, we omit MHD turbulence and simulate only mild eccentric waves that do not generally shock. In Paper II, we examine both MRI turbulence and stronger eccentric waves, and thus assess the relative importance of turbulent damping against wave non-linearity, and their relationship to r-mode excitation.

\subsection{Theoretical complications and uncertainties}\label{sec:comp}
Like all of the explanations offered to explain HFQPOs thus far, the discoseismic model suffers from several theoretical uncertainties related to the accretion flows in black hole X-ray binaries. In this section we mention the most salient complications, and make a case for the value of a continued investigation into r-mode excitation.

\subsubsection{Magnetic fields and MHD turbulence}\label{sec:mag}
First, magnetic fields and MHD turbulence are essential components of the hot ionised flows near a black hole, and cannot be neglected. Large-scale, smooth poloidal magnetic fields alter r-mode trapping if they are sufficiently strong \citep{FL09}. But this modification, which occurs due to an additional restoring force from magnetic tension, need not `destroy' the oscillations \citep{dew18}. Inhomogeneous radial distributions of vertical magnetic flux can in fact strengthen r-mode confinement (while enhancing frequencies). Regardless of the radial distribution of a poloidal field, \cite{kat17} argued that its effects on r-mode trapping may depend on the boundary condition assumed at the disc surface. Further, \cite{dew19} demonstrated that a strong toroidal magnetic field component (as is usually generated by the MRI) can reverse the negative effects of a vertical component.

Damping by radial inflow and turbulent fluctuations likely affects r-modes more strongly than ordered magnetic fields. While r-modes sometimes appear in hydrodynamic simulations \citep[see recent work by ][]{mis19}, the MHD simulations run by \cite{arr06} and \cite{ReM09} provided convincing evidence that the oscillations are not excited by MRI turbulence itself. The extent to which the MRI actively damps trapped inertial modes remains unclear, but analyses of the roles played by stresses in viscous disc models suggest that such damping is likely \citep[e.g.,][]{lat06}. We discuss the damping of disc oscillations by MHD turbulence in our companion paper.

\subsubsection{Disc geometry}
Another complication is the uncertainty regarding the disc's geometry during the very high (SPL) state. While large accretion rates appear to aid inward eccentricity propagation, the associated large radiation pressures may weaken r-mode trapping by inflating the vertical thickness of the disc. Indeed, classical $\alpha$-models of `bare' discs with accretion rates greater than 20-30\% of Eddington do not predict those discs to be geometrically thin \citep[e.g., ][]{lao89}. However, this vertical puffing is less pronounced in models that include the enhanced dissipation expected in the disc's coronal plasma \citep[e.g., ][]{Nay0}. Going beyond simple 1D models, the dynamics of radiation-pressure dominated accretion discs are not yet fully understood, since fully self-consistent 3D radiation-MHD simulations have only recently become feasible. \citet{jia14} found in radiation-MHD, shearing box simulations that a large fraction of accretion energy can be deposited in a self-consistently formed corona, contradicting 1D $\alpha$-models. More recently, \cite{Jia19} ran simulations of sub-Eddington accretion discs with opacities appropriate for AGN, and uncovered further deviations from the classic picture of thin-disc accretion. 

It has been suggested that in hard and intermediate (including SPL) states, X-ray binary accretion flows consist of a thin disc truncated at some radius outside the ISCO, with this radius only reaching the ISCO during `high-soft' states \citep[e.g.,][]{Done07}. Such truncation during the SPL state would preclude the excitation of trapped inertial oscillations close to the ISCO, as there would not be a disc, per se, at the r-mode trapping region. However, recent observations suggest that at least in some sources, the inner disc edge remains fixed near the ISCO during X-ray binary outbursts, the spatial extent of a hot corona changing over time instead \citep{kar19}. 

\subsubsection{Disc-corona connection}
One final issue of HFQPO phenomenology that remains to be addressed by all models is how to communicate dynamical oscillations present in the disc (or torus) to the high-energy, non-thermal emission associated with the corona. Poloidal magnetic fields provide one potential mechanism for this communication. Poloidal fields could stabilize r-modes against wave-breaking \citep{dew18}, and concurrently transfer their variability into time-dependent reconnection, dissipation, and ultimately emission in the coronal plasma \citep[see ][]{cab10}.

Clearly, HFQPOs in black hole X-ray binaries present a messy problem, with many underlying physical processes equally deserving of attention. In this and our companion paper we do not attempt to address all of these aspects, instead choosing to isolate the non-linear saturation of trapped inertial wave excitation via coupling with eccentric disc distortions, and the competition of this excitation with damping due to MHD turbulence.

\section{Simulation setup and tests}\label{sec:num}
\subsection{Equations}
The simulations presented in this paper have been run with the code RAMSES \citep{tey02,tey06}, which uses a finite volume, high-order Godunov method. In the absence of magnetic fields and explicit dissipation, the code solves the equations 
\begin{equation}
\label{eq:RAM1}
    \frac{\partial \rho }{\partial t}
    +\nabla\cdot(\rho {\bf u})= 0,
\end{equation}
\begin{equation}\label{eq:RAM2}
    \frac{\partial (\rho {\bf u})}{\partial t}
    +\nabla\cdot(
        \rho {\bf u u}
    )
    +\nabla P
    =-\rho\nabla\Phi,
\end{equation}
Here $\rho,{\bf u},P,$ and $\Phi$ are the mass density, fluid velocity, gas pressure and gravitational potential, respectively. For simplicity, we supplement Equations \eqref{eq:RAM1}-\eqref{eq:RAM2} with an isothermal equation of state $P=c_s^2\rho$, where $c_s$ is a purely constant sound speed. The potential $\Phi$ is taken to be the Paczynski-Wiita potential: 
\begin{equation}\label{PWpot}
    \Phi=\dfrac{-GM}{r-2r_g},
\end{equation}
where $G$ is the gravitational constant, $M$ is the mass of the central black hole and, for $c$ the speed of light, $r_g=GM/c^2$ is the characteristic gravitational radius.

\subsection{Numerical framework}
We use a version of RAMSES that solves Equations \eqref{eq:RAM1}-\eqref{eq:RAM2} on a uniform, cylindrical grid \citep{fau14}.
\footnote{This uniform grid version of RAMSES is freely available at \hyperlink{https://sourcesup.renater.fr/projects/dumses/}
{https://sourcesup.renater.fr/projects/dumses/}
} Under the cylindrical approximation \citep[e.g.,][]{arm98,haw01,sor12}, vertical gravity and density stratification are ignored. Our simulations are therefore radially global in the sense that they include curvature and non-local variations in the background equilibrium. We employ HLLD Riemann solvers in this and our companion paper, having confirmed that LLF and HLL solvers are more diffusive but provide similar results. Numerical expense precludes the simulation of the entire radial extent of the disc, and so we limit simulation domains to an annular region defined by inner and outer radii $r_0$ and $r_1.$ The vertical and azimuthal domains are $z\in[-H,H]$ and $\phi\in[0,2\pi),$ where $H=c_s/\Omega(r_\text{ISCO})$ is the isothermal scale height at the ISCO. 

\subsection{Physical parameters}\label{sec:phys}
In cylindrical coordinates $(r,\phi,z)$, the most basic equilibrium to consider is that of rotation dictated by the force balance
\begin{equation}\label{eq:1dEqm}
    r\Omega^2
    =\dfrac{1}{\rho}\dfrac{\textrm{d}P}{\textrm{d}r}
    +\dfrac{\textrm{d}\Phi}{\textrm{d}r}.
\end{equation}
Even without a cylindrical approximation, $\Omega=\Omega(r)$ is assured to be a function only of radius for a globally isothermal equation of state. The gravitational radius $r_g=GM/c^2$ defines the unit of length in the simulations. With $c=G=M=1,$ time units are in $\omega_g^{-1},$ where $\omega_g=c^3/(GM)$ is the gravitational frequency. In these units, the ISCO is located at $r_\text{ISCO}=6r_g,$ and the orbital period at this radius is given by $T_\text{orb}\approx 61.56.$ The most relevant parameter is the sound speed, $c_s$, which with an isothermal equation of state serves as a direct proxy for temperature and disc thickness. We consider values $c_s=\{0.01c,0.02c\}$, which correspond to aspect ratios $H/r\sim \{0.016,0.033\}$ at the ISCO. 

\subsection{Boundary conditions}\label{sec:BC}
In the 3D simulations described in Section \ref{sec:3D}, we have chosen an inner boundary $r_0=4r_g<r_\text{ISCO}$. Our reasons for placing $r_0$ within the ISCO are threefold: first, allowing material to flow through the ISCO is more physically realistic; second, this choice produces simulations more directly comparable to those of \cite{ReM09}; third, setting $r_0<r_\text{ISCO}$ significantly reduces the excitation of spiral density waves via the corotational instability considered by \cite{FL09} and \cite{FL13} (see Appendix \ref{sec:corot}). Note that we do discuss test simulations with the inner boundary placed at the ISCO itself ($r_0=6r_g$) in Appendix \ref{app:lintr} and Appendix \ref{sec:corot}. 

In our primary simulations, we use a `diode' outflow boundary condition at this inner boundary. In the ghost cells, this boundary condition matches density and vertical mass flux to the value in the closest cell of the active domain. We set the radial mass flux $\rho u_r$ to the value in the last cell in the active domain as long as this value is negative (i.e., flowing inward, out of the simulation domain); otherwise, it is set to zero. Finally, a perturbation from the background `pseudo-Keplerian' rotational velocity determined by the Paczynski-Wiita potential is calculated at the innermost active cell, and added to an extrapolation of this background rotational velocity in each ghost cell. 

We choose an outer radial boundary condition that allows us to produce an eccentric deformation in the simulation domain. At the outer radius $r_1$, we set $u_r=u_z=0$ and match $u_\phi$ to pseudo-Keplerian rotation. We then generate disc eccentricity solely by imposing a non-axisymmetric, precessing surface density profile in the ghost cells at the outer radial boundary. This produces a non-axisymmetric pressure gradient in the outer disc, which in turn forces the inward propagation of eccentricity. 

To calculate appropriate surface density profiles and retrograde precession frequencies to impose at the outer radial boundary, we first compute eccentric standing-wave solutions of the non-linear secular theory considered by \cite{bar16}. These non-linear eccentric eigenmodes achieve a maximum eccentricity (denoted $A_f$) within a specified radial domain, transition to circular streamlines at the radial boundaries \citep[cf., fig. 1 in ][]{bar16}, and precess in a retrograde fashion at a frequency $\omega_P<0$. While the velocity fields associated with the eccentric eigenmodes vanish at the circular radial boundaries, the density perturbation $\rho_E(r=r_1,\phi,t)$ is non-zero, possessing a periodic structure in $\phi$ that is \emph{nearly} sinusoidal at low $A_f$ (see Fig. \ref{fig:2Dstrms}, right).

These Newtonian eccentric disc solutions cannot simply be initialized in the domain of our \emph{relativistic} simulations; along with a different gravitational potential, our simulations use an inner boundary condition that favours travelling eccentric waves, not standing waves. But we find that setting the density in the ghost cells to $\rho(r=r_1,\phi,t)=\rho_E(r=r_1,\phi+\omega_Pt)$  organically induces a distorted, eccentric disc throughout our domain. In calculating the eccentric modes as described by \cite{bar16}, we take the ratio between inner and outer boundaries as $r_1/r_\text{ISCO}$, and scale units of length and time to match the code units of $r_g$ and $\omega_g^{-1}$. The precession frequencies are very small under this scaling, but we find that they must be imposed to produce a quasi-steady eccentric disc.

We characterize the non-axisymmetric density profiles imposed at $r_1$ by the maximum eccentricity found in the corresponding (Newtonian) eccentric mode, denoted as $A_f.$ We stress, however, that the maximum eccentricities induced in a given simulation are significantly lower than $A_f$. This discrepancy occurs in part because we only enforce density profiles calculated from the non-linear eigenmodes in the outer ghost cells, and the resulting perturbations in the active domain are smaller in amplitude. Even without this numerical effect, we would expect the different angular momentum distribution produced by the Paczynski-Wiita potential, as well as the the diode condition imposed at $r_0$, to alter the structure of the distortion. The diode condition allows the continued inward propagation of the travelling eccentric wave, instead of the reflection required to form an eccentric standing mode.

Periodic boundary conditions are imposed in azimuth. While periodic vertical boundary conditions are technically the appropriate choice for a cylindrical model, we find that in conjunction with the excitation of inertial waves in eccentric discs, such BCs lead to the formation of mean vertical flows that are constant in time (i.e., not oscillatory), and probably an artifact of the absence of vertical stratification (see Section \ref{sec:pUBC}). Therefore, unless otherwise stated, we focus on simulations with the vertical velocity set to zero in the ghost cells at $z=\pm H$ (periodic boundary conditions are still imposed upon the other fluid variables).

Aside from halting the growth of the mean vertical flows, this choice of vertical boundary condition primarily impacts on the phase of vertically structured oscillations; in particular, the r-modes in our simulations have vertical structure described essentially by sines for vertical velocity, and cosines for the rest of the fluid variables. The oscillations' horizontal velocity components therefore possess even symmetry with respect to the mid-plane, and might be identified most closely with $n=2$ r-modes in a fully stratified model. While it is true that our quasi-rigid vertical boundary condition excludes modes of odd parity, which in a stratified disc would bear the most physical relevance, any distinction based on mid-plane symmetry is not particularly meaningful in a cylindrical, unstratified framework.

\subsection{Initial conditions}\label{sec:IC}
To generate an initial condition for our three-dimensional simulations, we follow a three-step process. We first initialize 1D simulations with density set to a floor value within $r_\text{ISCO}$, and a constant background $\rho_0$ without. Equation \eqref{eq:1dEqm} determines the initial rotational velocity. Following a redistribution of angular momentum, these simulations produce a steady state in which the density falls off smoothly toward the ISCO, and mass continues to trickle through the inner boundary ($\dot{M}/M_{tot}\approx10^{-8}\omega_g$). 

\begin{figure}
    \centering
    \includegraphics[width=\columnwidth]{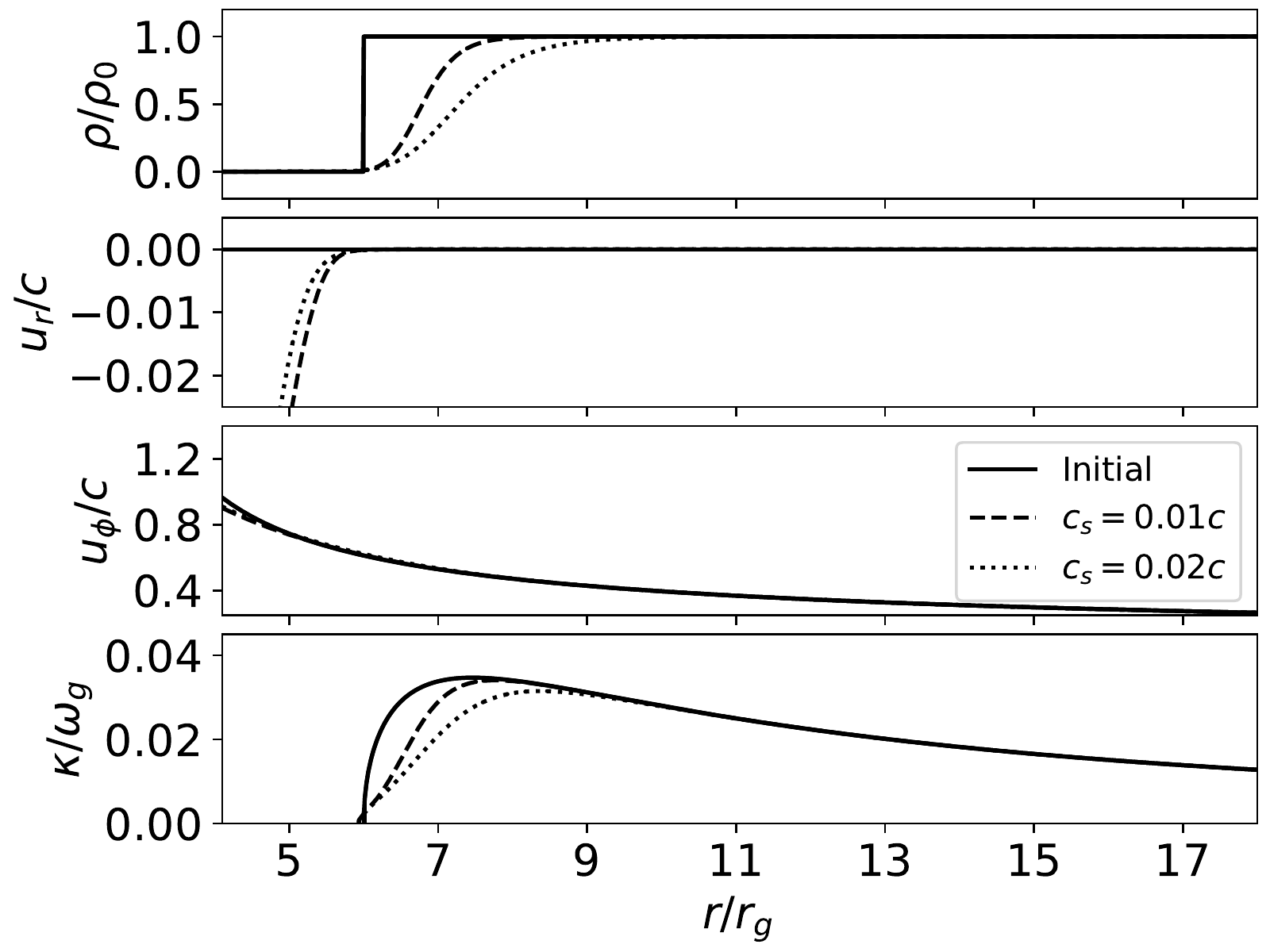}
    \caption{Radial profiles for the density, horizontal velocity components and horizontal epicyclic frequency calculated at $1000T_\text{orb}$ in 1D simulations run with $c_s=0.01c$ (dashed) and $c_s=0.02c$ (dotted). The solid lines describe the initial conditions, which relax to the final profiles plotted with dashed and dotted lines following an initial redistribution of angular momentum.}
    \label{fig:1dtran}
\end{figure}

The flow is transonic in that the inward radial velocity surpasses the sound speed in amplitude, but this occurs only in the inner, evacuated regions where the density becomes very small ($\sim10^{-4}\rho_0).$ The plots in Fig. \ref{fig:1dtran} illustrate the radial profiles of the density and horizontal velocities, along with the horizontal epicyclic frequency (calculated as $\kappa^2=2\Omega[2\Omega+r\partial_r\Omega]$) for simulations run with sound speeds $c_s=0.01c$ and $0.02c.$ The figure indicates that although deviations from the initial condition are small, they result in an outward shift of $\kappa$'s maximum. This shift is important to note before attempting to diagnose trapped inertial oscillations.  

\begin{figure}
    \centering
    \includegraphics[width=\columnwidth]{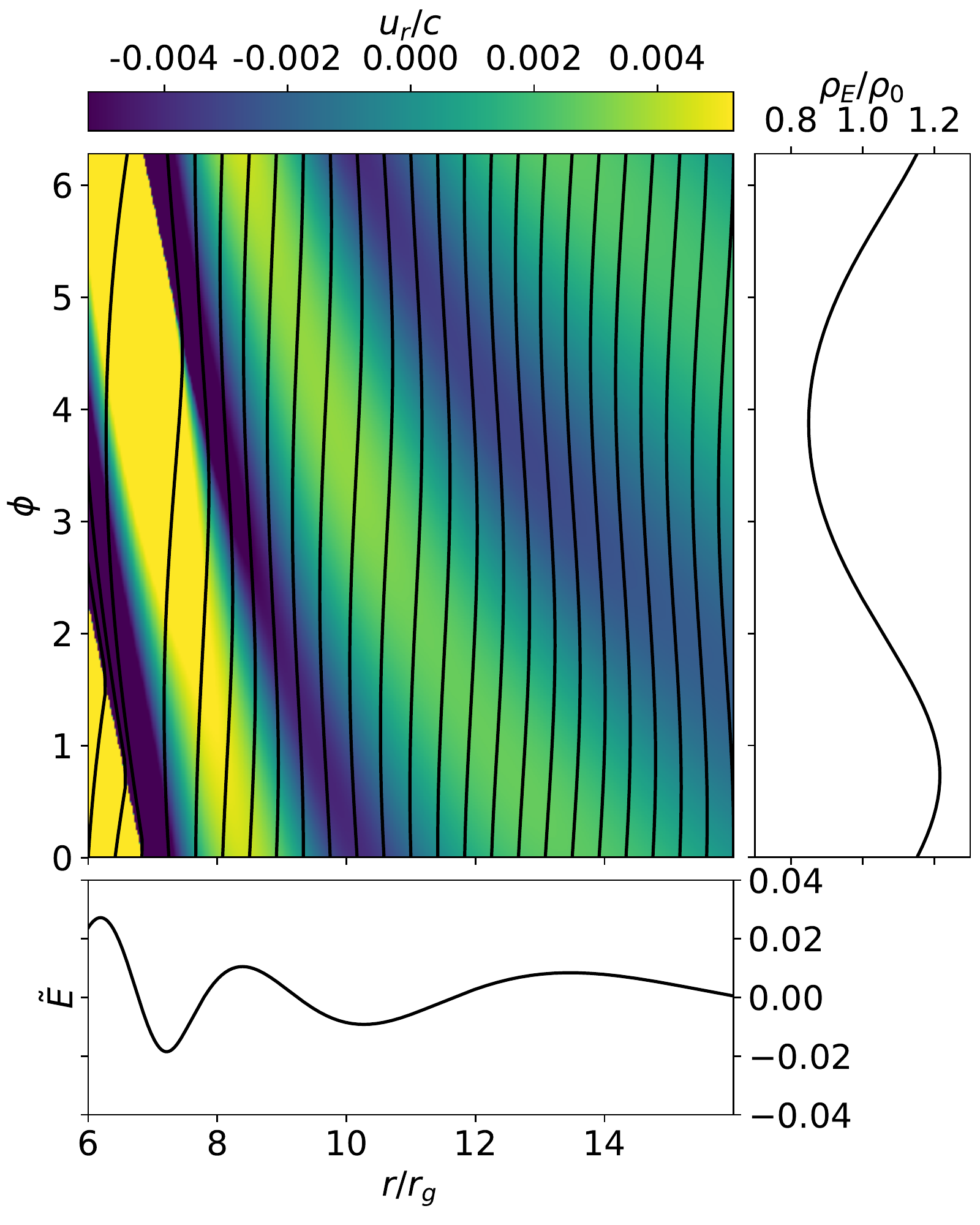}
    \caption{Top left: colour-plot showing radial velocity at the end of a 2D simulation run to generate an eccentric disc to use as an initial condition ($c_s=0.02c$, $A_f=0.1$). Top right: azimuthal profile of the non-axisymmetric density profile enforced in the outer radial ghost cells. Forced to precess very slowly in retrograde ($\omega_P\sim -4\times10^{-4}\omega_g$ in this run), such density profiles are solely responsible for the quasi-stationary eccentric structures in our simulations. The streamlines overlaid as black lines on the colour-plot indicate the resulting disc distortion, since circular streamlines would appear vertical on the unfolded polar plot. Bottom: linear approximation to the (imaginary part of) the complex eccentricity, extracted from the simulation snapshot above.}
    \label{fig:2Dstrms}
\end{figure}

After integrating these 1D simulations for $1000T_\text{orb}$, we copy the resulting radial profiles for each fluid variable at each azimuthal grid cell to initialize 2D simulations in $(r,\phi)$ aimed at producing an eccentric, relativistic disc. The outer radial boundary condition described in Section \ref{sec:IC} then produces discs with non-circular streamlines. Without significant reflection at $r_\text{ISCO}$, the eccentricity continually propagates through the inner boundary as an eccentric `travelling wave', similar to those calculated by \cite{fer09}. The resulting quasi-steady state achieved after $100T_\text{orb}$ is that of a \emph{twisted} eccentric disc, with streamlines composed of concentric ellipses with semi-major axes angled with respect to one another. Fig. \ref{fig:2Dstrms} illustrates the eccentric end-state of a 2D simulation with $c_s=0.02c$. The top left colour-plot shows radial velocity overlaid by distorted streamlines, while the top right plot shows the imposed azimuthal variation of density in the ghost cells required to generate the eccentric structure. The bottom plot shows a linear approximation to the imaginary part of the complex eccentricity, $\tilde{E}=\int \langle u_r\rangle_z\cos\phi\text{d}\phi/(\pi\langle u_\phi\rangle_{\phi,z})\sim -e\sin\varpi$. The radial structure
in $\tilde{E}$ resembles qualitatively the undamped profiles of complex eccentricity computed by \cite{fer09}, in that the wave grows and oscillates on shorter radial scales as it approaches the ISCO.

Sufficiently large forcing amplitudes at the outer boundary produce strong, non-axisymmetric variations in both density and velocity in the evacuated regions within the ISCO (see the left edge of Fig. \ref{fig:2Dstrms}). However, for forcing amplitudes less than or equal to those implemented here, the streamlines remain smooth in the expected trapping region close to the maximum in $\kappa$ (located at $r\gtrsim 8r_g$ for the simulation shown in Fig. \ref{fig:2Dstrms}). We find no evidence that the r-mode excitation could be affected by the flow within the ISCO. However, these strong variations may play a role in damping small-scale fluctuations that grow in the inner (unstable) regions in our control simulations of circular discs, but which disappear with even a very small eccentric forcing.  

We use the end states of the 2D simulations as the initial condition for our 3D simulations. The two-dimensional, eccentric disc flow is copied at each vertical grid cell, and superimposed with white noise velocity perturbations at amplitudes $<0.01c_s$.

\subsection{Diagnostics}
The most basic diagnostic utilized is a volume average, defined for a quantity $X$ by
\begin{equation}
    \langle X\rangle_V = \dfrac{\int_V X dV}{\int_VdV}.
\end{equation}
We apply similar definitions for azimuthal and vertical averages, denoted as $\langle X\rangle_{\phi}$ and $\langle X\rangle_z$. Additionally, we also consider volume averages within the annular domain $D$ defined by $r/r_g\in[7,9]$, which circumscribes the expected trapping region. 

Lastly, timing analyses provide an important window into the nature of the oscillations excited. We therefore consider power spectral densities (PSDs) calculated from time-series data. Defined as $P(\omega)\propto|\mathcal{F}(f)|^2,$ where $\mathcal{F}$ is the Fourier transform in time, the PSD describes the power in a given frequency of oscillation exhibited by the signal $f(t)$. We primarily present power spectral density as a function of both frequency and radius. In such cases, time series have simply been collected at fixed radial grid-points independently before being used to calculate the PSD.

\subsection{Test simulations}\label{sec:2D}
In addition to the 3D simulations discussed in Section \ref{sec:3D}, in Appendix \ref{sec:tests}  we also present the results of preliminary test simulations. In Appendix \ref{app:lintr}, we describe the evolution of linear r-modes inserted by hand into `2.5D' (axisymmetric) simulations run with resolutions and boundary conditions matching those of our simulations of eccentric discs. These simulations confirm that RAMSES is capable of capturing the purely oscillatory behavior of r-modes in isolation, and provide a point of comparison for the modes discussed in Section \ref{sec:3D}. They further reveal that the oscillations give rise to weak `zonal flows' (vertically homogeneous modifications to the background equilibrium flow) through a local redistribution of angular momentum. These zonal flows can alter the trapping region, and may be responsible for small changes in r-mode frequency over time. 

In Appendix \ref{sec:corot}, we validate the performance of our code in the non-linear regime by reproducing results from \cite{FL13}. In addition to points of comparison and insight into our 3D runs, both suites of simulations provide motivation for our choice to allow material to flow through an inner boundary placed within the ISCO.

\section{Results}\label{sec:3D}
In this section we present our primary 3D simulations of eccentric, relativistic discs. We demonstrate that trapped inertial oscillations can be excited by imposed eccentric structures, and then describe the modes' non-linear saturation.

Table \ref{tab:trExHsim} provides a summary of the runs considered.  All have been initialized with the end states of 2D planar simulations, such as the one pictured in Fig. \ref{fig:2Dstrms}, superimposed with white noise velocity perturbations. Each simulation covers a domain $[r_0,r_1]\times[\phi_0,\phi_1]\times[z_0,z_1]=[4r_g,18r_g]\times[0,2\pi)\times[-H,H]$, with $H$ the scale height at the ISCO. A resolution of $512\times512\times32$ at $c_s=0.02c$ then yields grid cells with aspect ratio $\text{d}r$:$r\text{d}\phi$:$\text{d}z\sim $2:6:1 at the ISCO. The cells are elongated azimuthally, but both the twisted eccentric modes and the oscillations with which they couple involve only low azimuthal wavenumbers, which should be more than adequately resolved. Indeed, doubling resolution in each (and every) spatial direction reveals growth rate convergence for $r\text{d}\phi/\text{d}r\lesssim 3$ (see Appendix \ref{app:conv}). Simulation f10c2p is analogous to f10c2 but has been run with purely periodic vertical boundary conditions for comparison.

\begin{table}
\centering
\caption{Table summarizing 3D, hydrodynamic, pseudo-Newtonian simulations of relativistic discs, performed on the domain $[4r_g,18r_g]\times[0,2\pi)\times[-H,H]$. From left to right, the table lists the (i) simulation label, (ii) amplitude for eccentricity forcing, (iii) approximate eccentricity near the trapping region, (iv) sound speed, (v) resolution, and (vii) estimated r-mode growth rate. All simulations have a run-time of $T_\text{max}=5000\omega_g^{-1}.$}\label{tab:trExHsim}
\begin{tabular}{lccccr} 
    \hline
    Label    &
    $A_f$    &
    $e_\text{tr}$ & 
    $c_s/c$  &
    $N_r\times N_{\phi}\times N_z$ &
    $s/\omega_g$ \\
    \hline
    ctrl2 & $0$     & $0$     & $0.02$ & $512\times 512\times 32$ & - - \\
    f02c2 & $0.025$ & $0.003$ & $0.02$ & $512\times 512\times 32$ & - - \\
    f05c2 & $0.05$  & $0.006$ & $0.02$ & $512\times 512\times 32$ & $4.3\times 10^{-3}$\\
    f07c2 & $0.075$ & $0.009$ & $0.02$ & $512\times 512\times 32$ & $4.9\times 10^{-3}$\\
    f10c2 & $0.10$  & $0.013$ & $0.02$ & $512\times 512\times 32$ & $5.8\times 10^{-3}$\\
    \hline
    f10c2p & $0.10$ & $0.013$ & $0.02$ & $512\times 512\times 32$ & $5.3\times 10^{-3}$\\
  \hline
    ctrl1 & $0$     & $0$     & $0.01$ & $1024\times1024\times 32$ & - - \\
    f05c1 & $0.05$  & $0.002$ & $0.01$ & $1024\times1024\times 32$ & $1.0\times 10^{-3}$\\
    f10c1 & $0.10$  & $0.006$ & $0.01$ & $1024\times1024\times 32$ & $3.3\times 10^{-3}$\\
    \hline
 \end{tabular}
\end{table}

The table also lists parameters describing r-mode excitation. $A_f$ describes the maximum eccentricity of the eigenmode used to force the outer boundary. We note again that due to the continual propagation through $r_0,$ the resulting eccentricity within the disc is much lower than $A_f$. This is indicated by the values listed for $e_\text{tr}$, which gives an approximate measure of the disc eccentricity in the expected r-mode trapping region. This is calculated from $e_\text{tr}\sim\max|u_r|/\langle u_\phi\rangle_{\phi,z}$ at the radius of maximum $\kappa$ shown in Fig. \ref{fig:1dtran}. The values of eccentricity measured for a given forcing amplitude are smaller for lower sound speeds, due to a tighter `winding' of the twisted eccentric mode close to the ISCO. The values of $e_\text{tr}$ listed in Table \ref{tab:trExHsim} are relatively low, but prove sufficient to excite trapped inertial waves: the final column lists estimates of the oscillations' growth rates, calculated as described in the following section.

\subsection{Volume-averaged quantities}
\begin{figure*}
    \centering
    \includegraphics[width=\columnwidth]{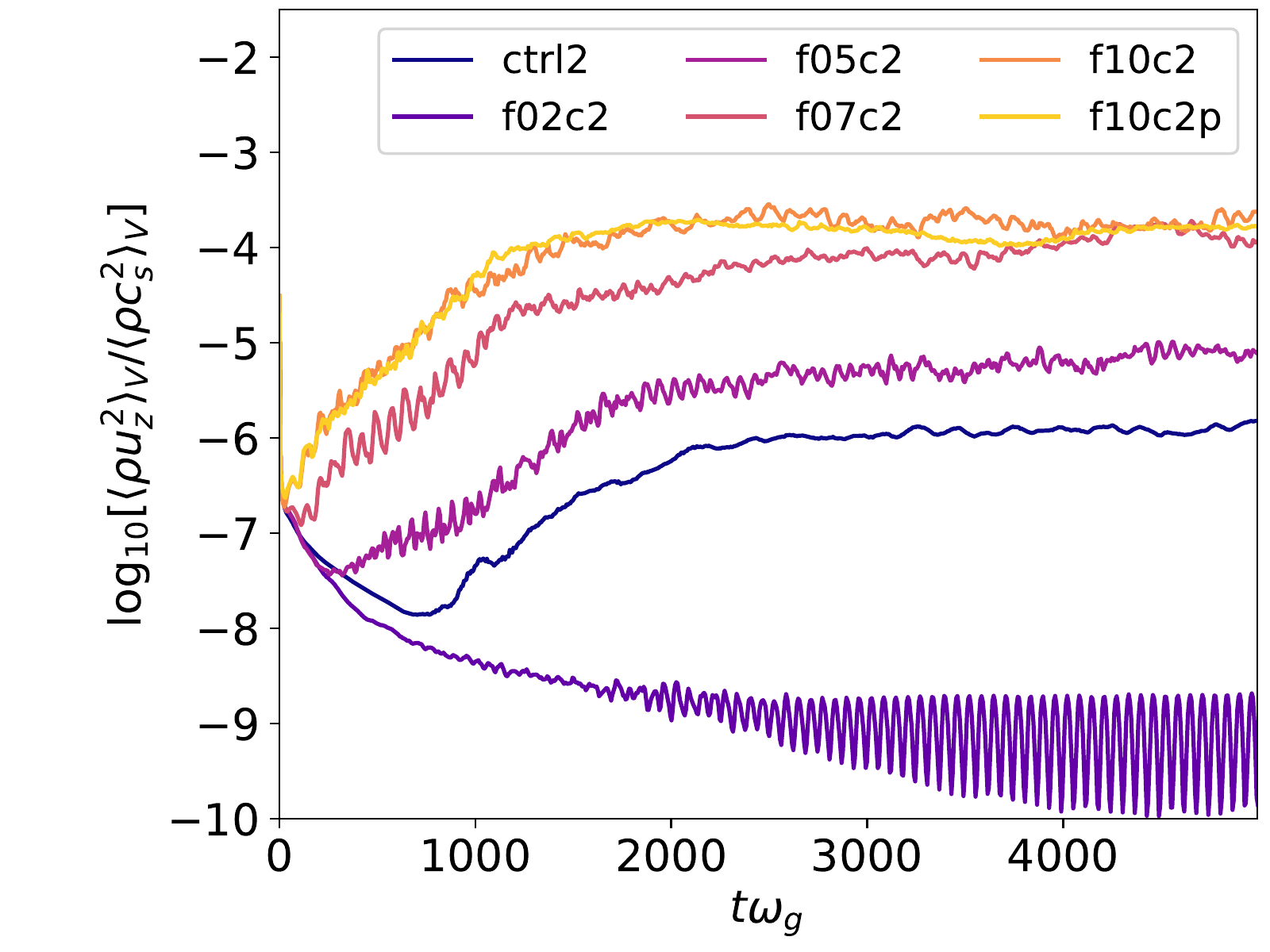}
    \includegraphics[width=\columnwidth]{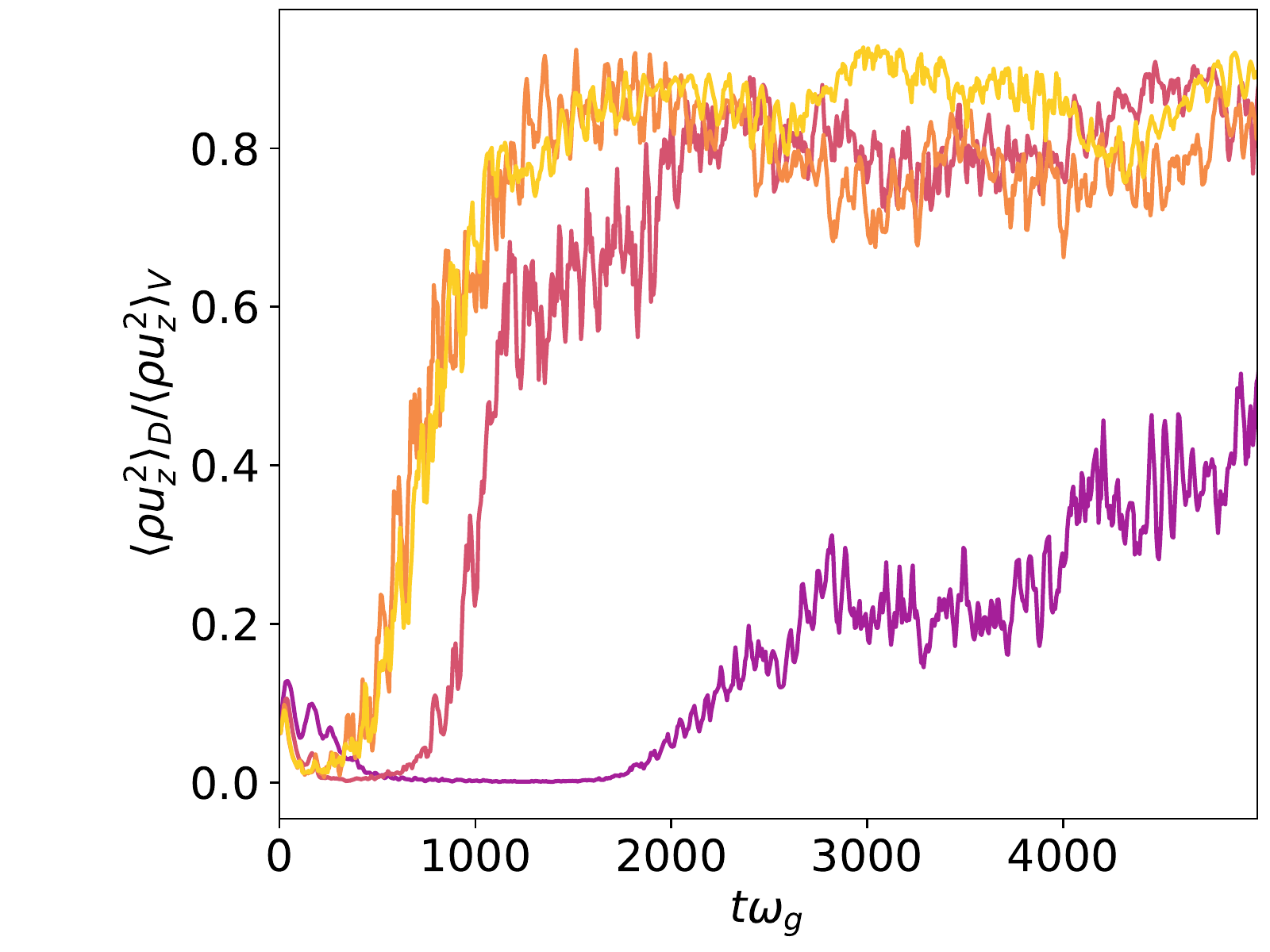}
    \includegraphics[width=\columnwidth]{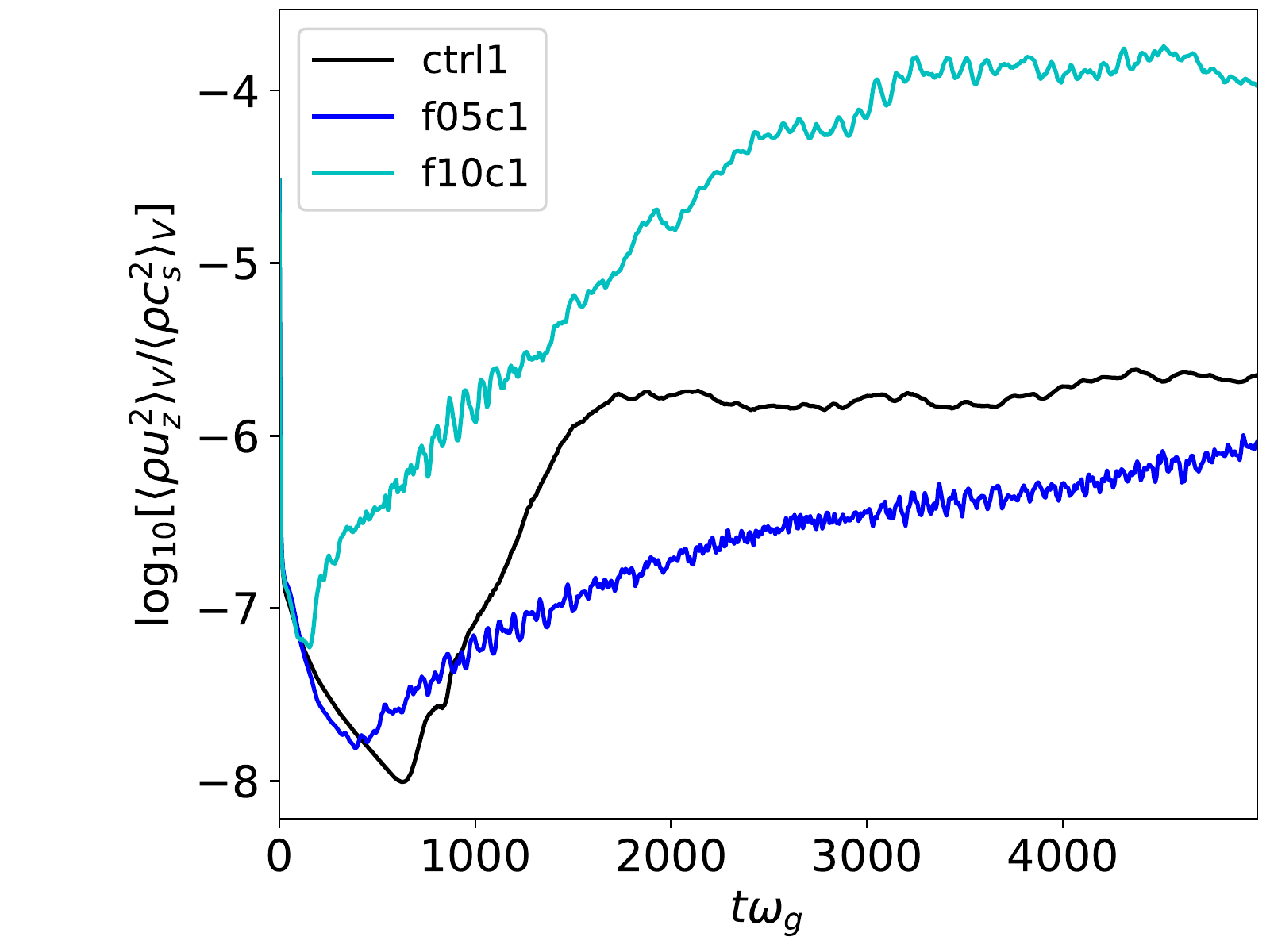}
    \includegraphics[width=\columnwidth]{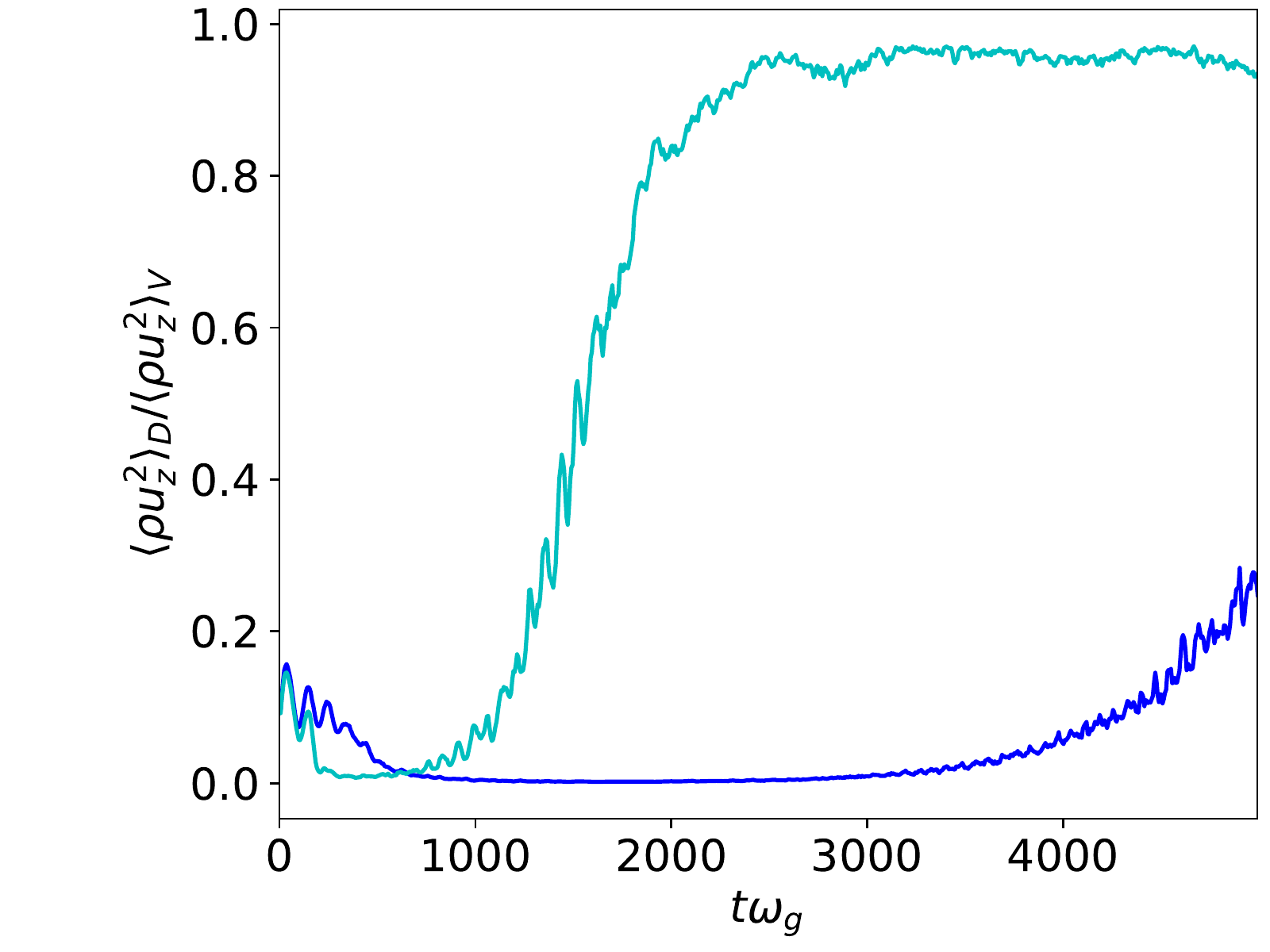}
    \caption{Left: Volume-averaged vertical kinetic energy density for the simulations listed in Table \ref{tab:trExHsim} with $c_s=0.02c$ (top) and $c_s=0.01c$ (bottom), normalised by the kinetic energy associated with sonic motions. Right: fraction of vertical kinetic energy contained within the annular domain D defined by $r/r_g\in[7,9].$ The growth in vertical kinetic energy follows the excitation of trapped inertial waves.} \label{fig:htrKE}
\end{figure*}

Fig. \ref{fig:htrKE} illustrates the growth of the volume-averaged vertical kinetic energy density in the simulations listed in Table \ref{tab:trExHsim}. This can be used to diagnose vertical oscillations \citep{ReM09}. Note that the cylindrical framework does not capture the `breathing' that occurs in fully 3D, density stratified models of eccentric discs, which arises from the variation in vertical gravity around elliptical streamlines \citep{ogi01}. The vertical kinetic energy can therefore be dissociated from the 2D eccentric structures considered in these simulations, and directly associated with the modes they excite. 

The left-hand panels show vertical kinetic energy density volume-averaged over the entire domain in simulations run with $c_s=0.02c$ (top) and $c_s=0.01c$ (bottom), and normalized by the volume-averaged pressure.  This is equivalent to the vertical kinetic energy of oscillations in the disc normalized by the kinetic energy associated with sonic motion, so Fig. \ref{fig:htrKE} (top left) implies a saturated rms $|v_z^2|^{1/2} \sim 0.015c_s$ for simulation f10c2. This value is low, but is tied directly to the small disc eccentricities (of only $e_\text{tr}\lesssim0.013$ near the trapping region) that excite them.

Meanwhile, the right-hand panels show the fraction of vertical kinetic energy contained in the annular domain $r/r_g\in[7,9].$ Figs. \ref{fig:htrKE} (right) show that the enhancement of vertical oscillations' amplitudes indicated by Figs. \ref{fig:htrKE} (left) takes place within the trapping region expected for r-modes. The growth seen in simulation f10c2 is mirrored by its counterpart with periodic vertical boundary conditions, f10c2p. 

Notably, the circular disc simulations without forced eccentricity (ctrl1 and ctrl2) do show a small growth in vertical kinetic energy. This is due to the activity of high-$m$ fluctuations close to and within the ISCO. Regardless of their origin, these fluctuations contribute an averaged vertical kinetic energy that is dwarfed by that exhibited in the non-circular disc simulations, and disappears with even a very small amplitude eccentricity forcing (see simulation f02c2). As mentioned in Section \ref{sec:2D}, this may be due to a damping of the fluctuations by strong density variations associated with streamline intersection in the inner evacuated regions.

Because they do not exhibit r-mode excitation, we exclude simulations ctrl1, ctrl2 and f02c2 from the right-hand panels in Fig. \ref{fig:htrKE}, and from r-mode growth rate estimations. We calculate the growth rates from our measurements of vertical kinetic energy density averaged over the domain D: assuming $\langle\rho v_z^2\rangle_D\sim \langle\rho_0\delta v_z^2\rangle_D,$ we fit exponentials of the form $\exp[2st]$ to find the growth rates listed in Table \ref{tab:trExHsim}. We do not observe the same quadratic scaling of growth rate with eccentricity found by \cite{fer08}, but we note that there are significant differences between our simulation set-up and their calculations. Importantly, the twisted, travelling eccentric modes used in our simulations possess a markedly different structure
to the $m=1$ standing modes considered by \cite{fer08}, since they have a phase that varies with radius. Eccentricity gradients play a role in r-mode excitation, and will also differ between their disc deformations and ours.

\subsection{Mode characterization}\label{sec:mods}
\begin{figure*}
    \centering
    \includegraphics[width=\textwidth]{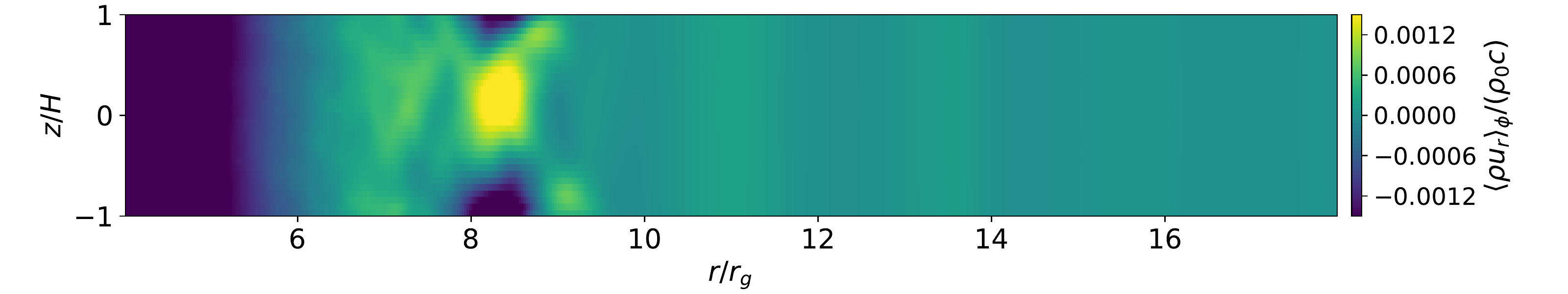}
    \caption{Meridional ($r-z$) colour-plot illustrating the trapped inertial wave in simulation f10c2, constructed by period-folding over the interval $1500<t\omega_g<2000$. Faint vertical columns at larger radii illustrate the outward propagation inertial-acoustic waves excited by the r-mode, also visible as diagonal streaks in the spacetime diagram given in Fig. \ref{fig:mflpr_spc} (second to bottom).}
    \label{fig:Af1c02r18_composite}
\end{figure*}

Fig. \ref{fig:Af1c02r18_composite} shows a visualization of the trapped inertial mode excited in simulation f10c2, constructed by period-folding azimuthally averaged radial mass flux over the interval $1500<t\omega_g<2000.$ The plot shows an inertial mode with $k_z\sim\pi/H$, trapped near $r\sim 8r_g$. Throughout the remainder of this section, we explore the timing properties of such trapped modes, and other oscillations excited in our simulations of eccentric discs.

\begin{figure*}
    \centering
    \includegraphics[width=\textwidth]{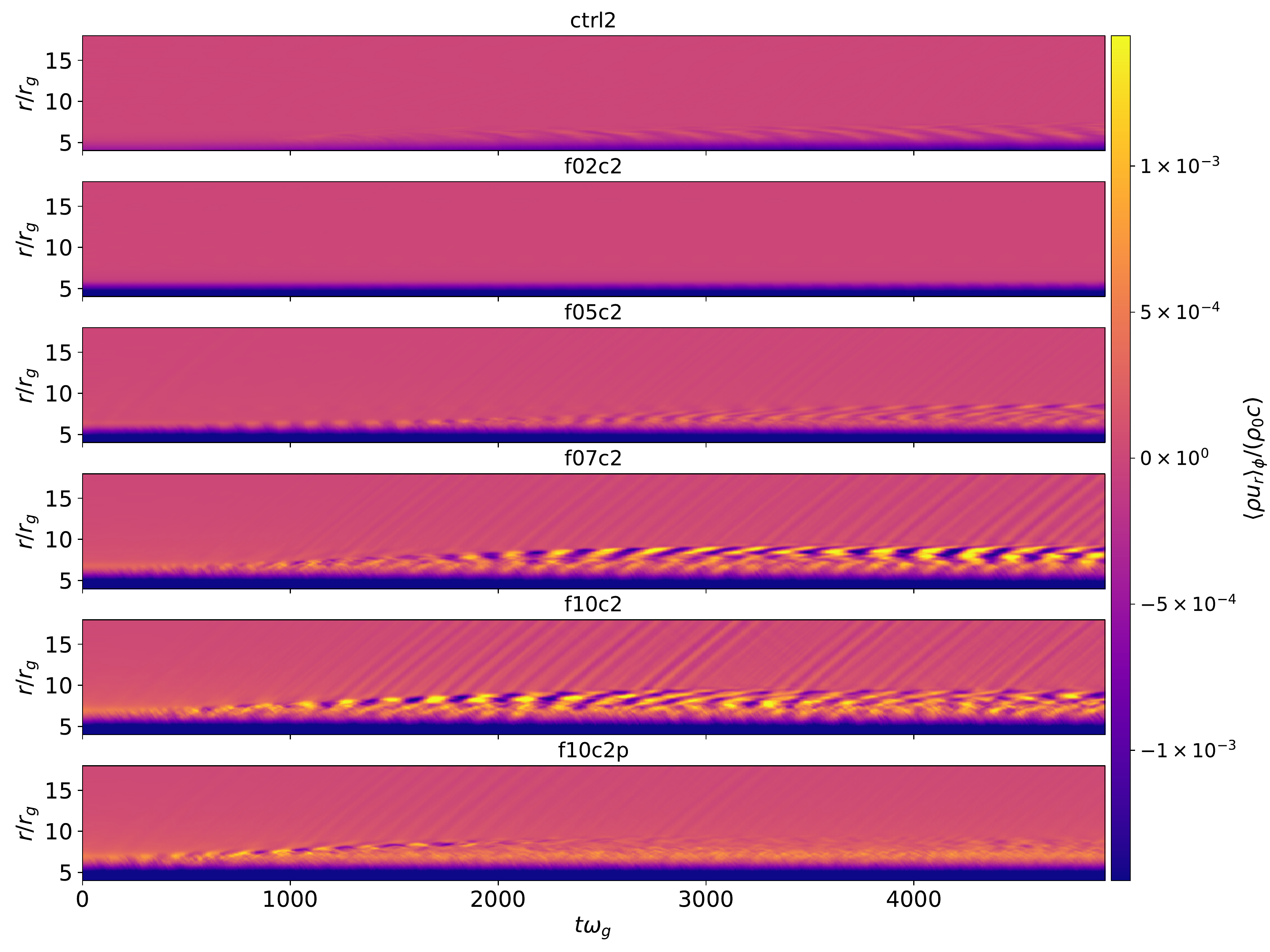}
    \caption{Space-time diagrams showing mid-plane, azimuthally averaged, radial profiles of radial mass flux, $\langle \rho u_r\rangle_{\phi}(r,z=0,t)$, for the simulations with $c_s=0.02c$ listed in Table \ref{tab:trExHsim}. The periodic fluctuations near $r\sim8r_g$ illustrate a trapped r-mode, while the diagonal lines at larger radii show outwardly propagating f-modes.}
    \label{fig:mflpr_spc}
\end{figure*}

\begin{figure*}
    \centering
    \includegraphics[width=\textwidth]{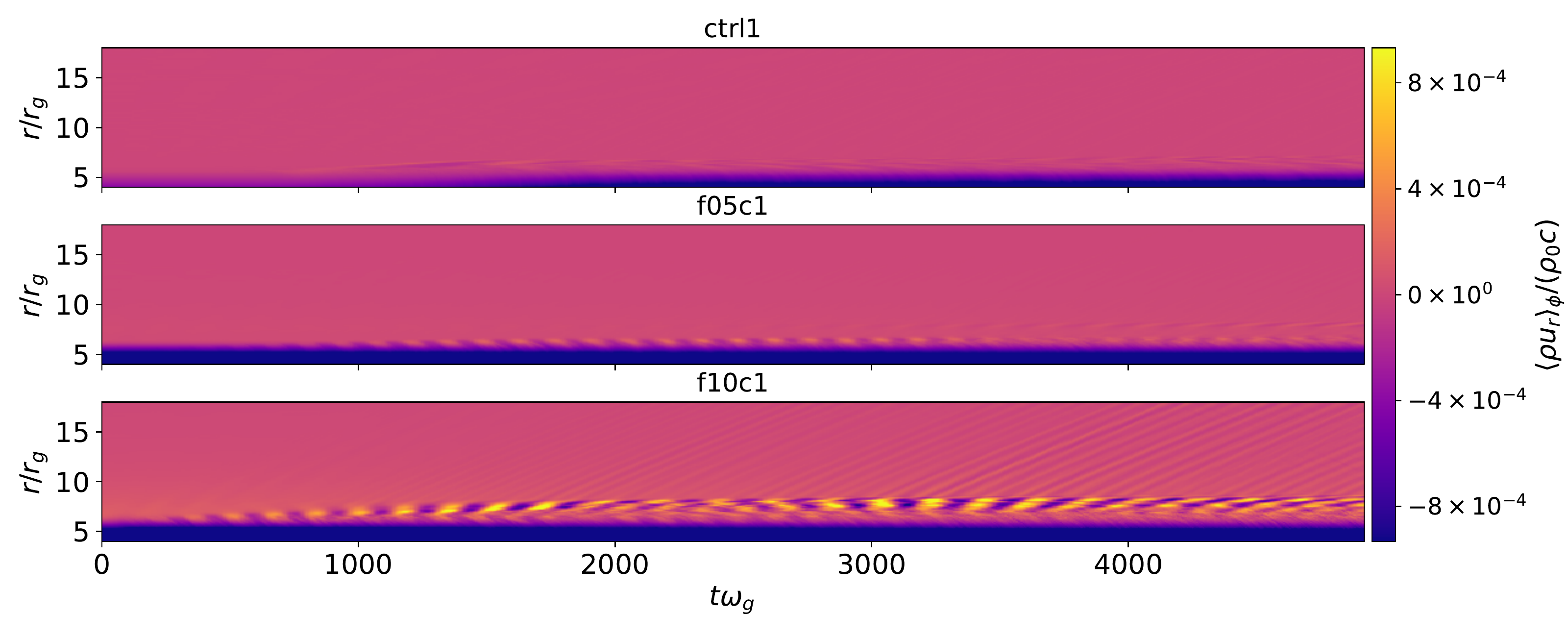}
    \caption{Same as in Fig. \ref{fig:mflpr_spc}, but for simulations run with $c_s=0.01c.$}
    \label{fig:mflpr_spc1}
\end{figure*}

The space-time diagrams shown in Figs. \ref{fig:mflpr_spc} and \ref{fig:mflpr_spc1} show mid-plane, azimuthally averaged profiles of radial mass flux over time for simulations run with $c_s=0.02c$ and $0.01c$, respectively. Along with Figs. \ref{fig:htrKE} (right), these space-time diagrams clearly reveal the emergence of oscillatory activity localised to the expected trapping region. The periodic nature of the growth in kinetic energy shown in Fig. \ref{fig:htrKE} is particularly well exhibited by the space-time diagrams for runs f07c2, f10c2 and f10c1. 

\begin{figure*}
    \centering
    \includegraphics[width=\columnwidth]{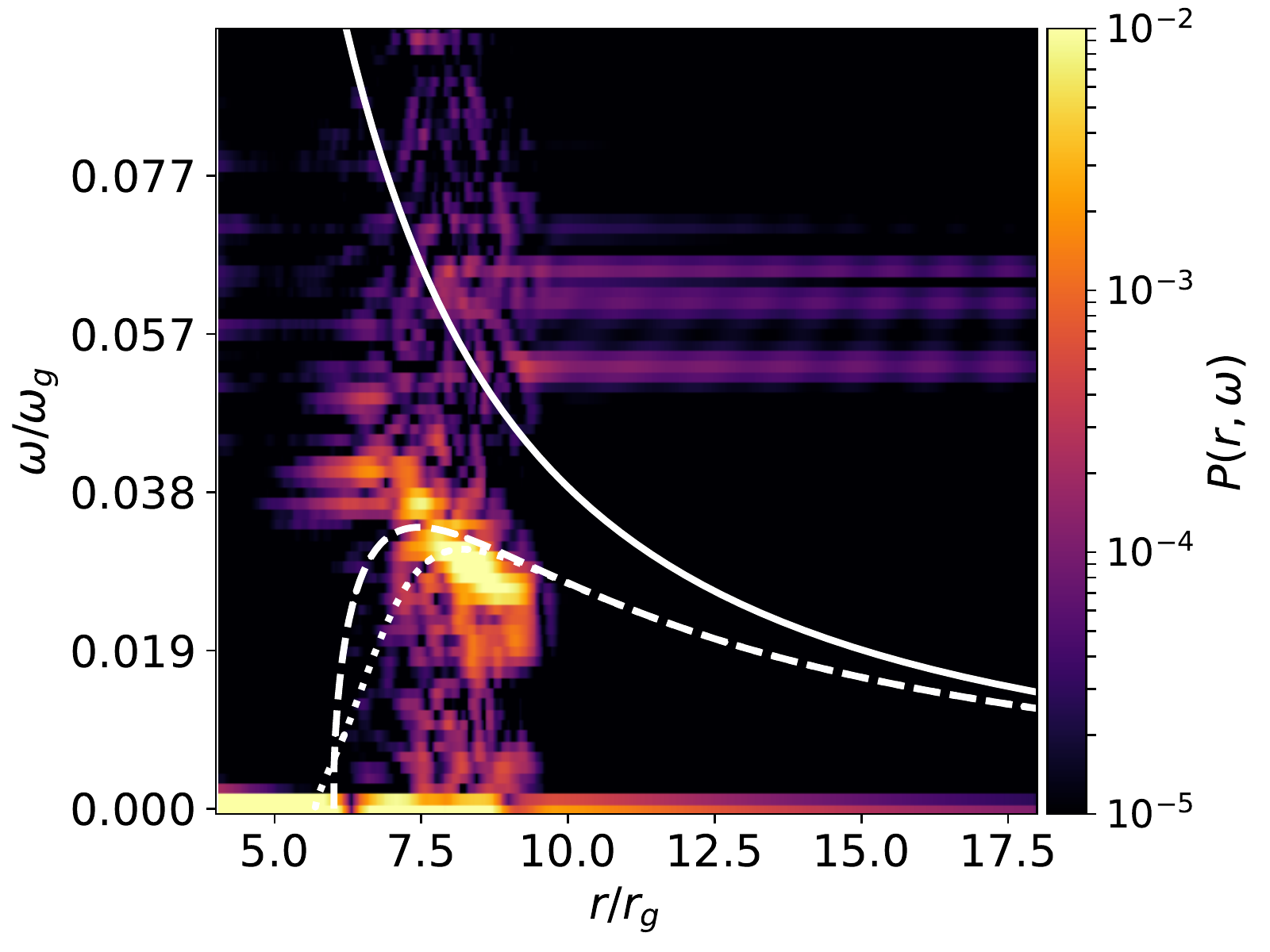}
    \includegraphics[width=\columnwidth]{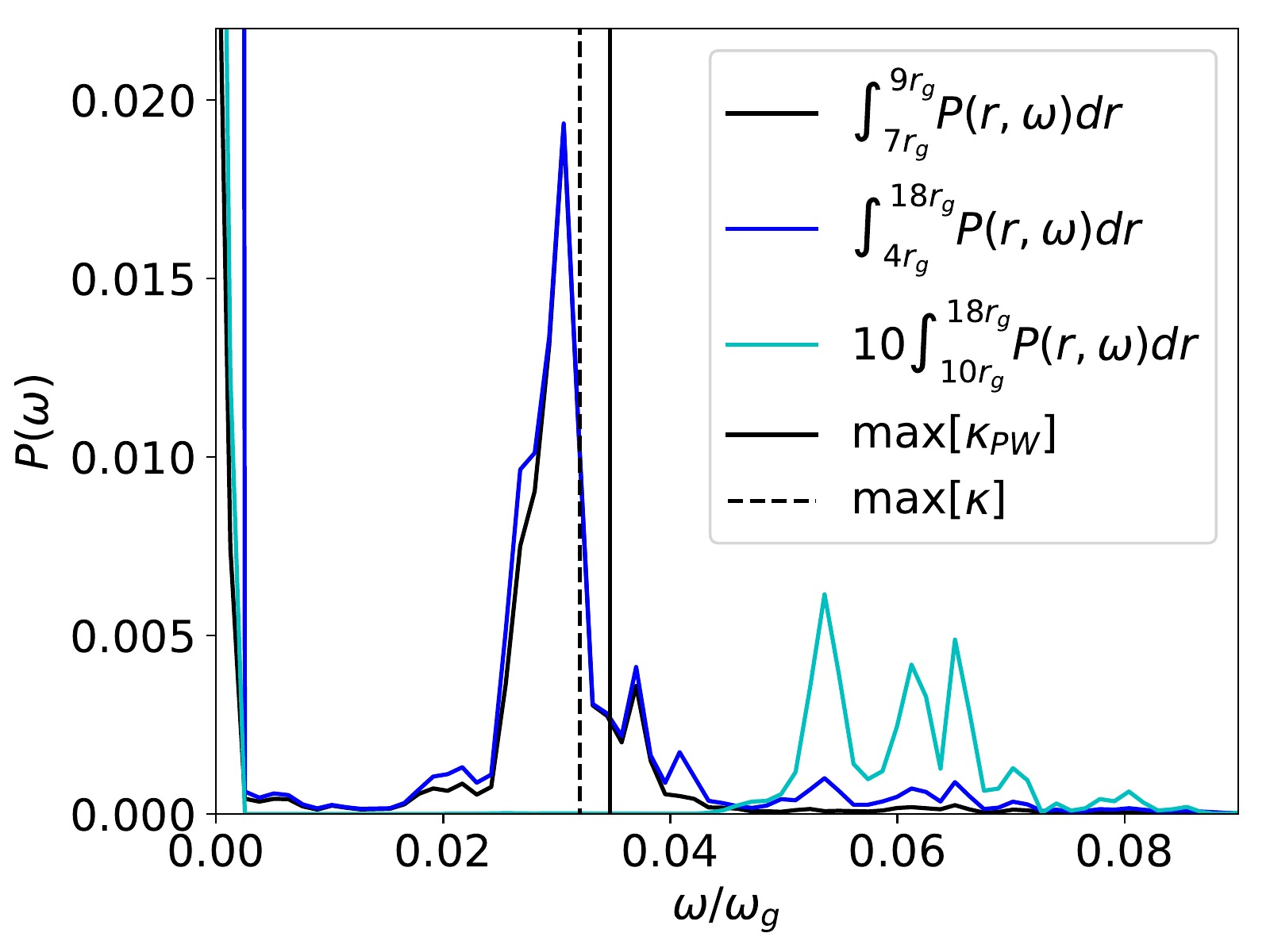}
    \caption{Left: Arbitrarily normalized power spectral density $P(r,\omega)$, calculated using radial profiles of $\langle\rho u_r\rangle_\phi(r,z=0,t)$ from simulation f10c2 (see Fig. \ref{fig:mflpr_spc}, second to bottom). The solid white line shows the analytical profile for the orbital frequency, while the dashed and dotted lines show the profiles for the horizontal epicyclic frequency calculated from a Paczynski-Wiita potential and from the 1D simulation run to establish an initial condition (resp.). Right: $P(\omega)$ integrated over various radial domains. The black and blue lines (integrated over the trapping region and the entire disc, resp.) show a dominant peak at the expected r-mode frequency, while the (amplified) cyan line shows the inertial-acoustic modes propagating in the outer disc.}
    \label{fig:Af1c02r18_mflp_psd}
\end{figure*}

Qualitatively, the frequencies appear consistent with predictions of $\omega/\omega_g\sim\max[\kappa_\text{PW}]\sim 0.035$ (or periods of $\sim181\omega_g^{-1}$), where $\kappa_\text{PW}$ is the horizontal epicyclic frequency determined solely by a Paczynski-Wiita potential. We confirm this quantitatively by computing the power spectral density $P(r,\omega)$, using the temporal data for $\langle\rho u_r\rangle_\phi$ shown in Figs. \ref{fig:mflpr_spc} and \ref{fig:mflpr_spc1}. The heatmap in Fig. \ref{fig:Af1c02r18_mflp_psd} (left) shows arbitrarily normalized power for simulation f10c2. The peaks at very low frequency at all radii correspond to a secular change in the background flow due to the precession of the twisted eccentric disc. Meanwhile, the strong peak at frequencies $\omega\sim 0.035\omega_g$ and radii $r\sim 8r_g$ indicates the growth of a trapped inertial wave. The peak is slightly offset from the analytical maximum for $\kappa_{PW}$, the epicyclic frequency calculated from a Paczynski-Wiita potential (plotted with a white dashed line), because of the modification to $\kappa$ that occurs in the 1D simulations run to establish an initial condition. The profile for $\kappa$ illustrated in Fig. \ref{fig:1dtran} is plotted with a dotted white line, and its peak aligns with the peak in power. The rotation profile calculated analytically from a Paczynski-Wiita potential is also plotted with a solid white line. 

The peak in power corresponding to the trapped inertial wave in Fig. \ref{fig:Af1c02r18_mflp_psd} (left) is surrounded by broadband noise. This blurring of the signal may be related to an outward shift of the r-modes' localization during the oscillations' saturation. This shift is most apparent in the space-time diagrams for simulations f07c2, f10c2 and f10c1 shown in Figs. \ref{fig:mflpr_spc} and \ref{fig:mflpr_spc1}. The black and dark blue lines in Fig. \ref{fig:Af1c02r18_mflp_psd} (right) show PSDs integrated over the radial domain $[7r_g,9r_g]$ and the entire disc (resp.), illustrating the dominance of the trapped inertial mode over the background noise. The black dashed and dotted lines correspond to the maxima in the analytical and actual profiles for the horizontal epicyclic frequency, respectively.

\begin{figure*}
    \centering
    \includegraphics[width=\textwidth]{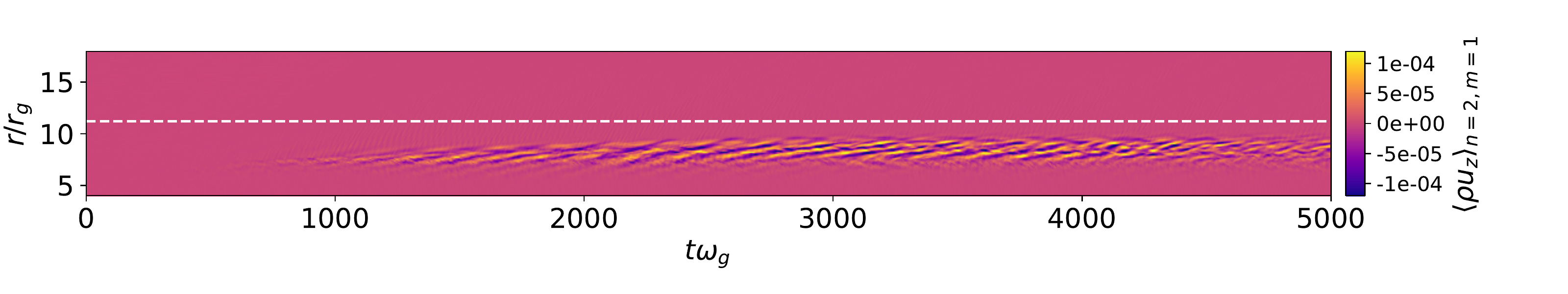}
    \caption{Spacetime diagram showing radial profiles of the average $\langle\sin(\pi z/H)\sin\phi \rho u_z\rangle_{\phi,z}$ over time. This average isolates variability associated with the intermediate, non-axisymmetric inertial wave that facilitates interaction between the axisymmetric ($m=0$) r-mode and the eccentric disc distortion. In agreement with semi-analytic calculations, the spacetime diagram illustrates a short-wavelength wave that propagates toward its corotation radius (white dashed line).}  
    \label{fig:n2m1spc}
\end{figure*}

While Figs. \ref{fig:mflpr_spc}-\ref{fig:Af1c02r18_mflp_psd} clearly indicate r-mode excitation, the signal is blurred. Test simulations seeded with linear normal mode solutions (see Appendix \ref{app:lintr}) indicate that this dynamical variation issues at least in part from the mode's back reaction on the underlying orbital motion. As demonstrated in Appendix \ref{app:lintr}, trapped inertial waves redistribute angular momentum locally, causing deviations from the orbital equilibrium, possibly in the form of zonal flows. If sufficiently strong, this redistribution alters the properties of the epicyclic frequency, in turn influencing the location and shape of the trapping region --- and hence the location and frequency of the trapped wave itself. We do not observe quasi-steady zonal flows directly in our 3D simulations, as seen in the local boxes of \citet{wei18}; it may be that the system is too dynamic for them to appear unambiguously. But the test simulations described in Appendix \ref{app:lintr} strongly suggest that trapped inertial waves will reconfigure the background flow as they reach non-linear saturation. The interplay between a trapped wave, its trapping region, and its time-varying redistribution of angular momentum would induce variation in the oscillation frequency, and might provide a partial explanation for HFQPOs' observed low quality factors.

In order to verify that the trapped mode excitation indicated by Figs. \ref{fig:mflpr_spc}-\ref{fig:Af1c02r18_mflp_psd} proceeds via the three-wave coupling described in Section \ref{sec:bgEx}, we seek to identify the intermediate, non-axisymmetric (i.e., $m=1$) inertial wave that facilitates interaction between the axisymmetric r-mode and the eccentric distortion. To identify variability associated with this intermediate wave, we consider the average
$\langle\rho u_z\rangle_{n=2,m=1}
\equiv\langle\sin(\pi z/H)\sin\phi\rho u_z\rangle_{\phi,z}$. This average isolates the $m=1$ component of the vertical mass flux, while also removing the background flow associated with the eccentric disc.

The spacetime diagram in Fig.~\ref{fig:n2m1spc} shows radial profiles of $\langle\rho u_z\rangle_{n=2,m=1}$ over time, revealing an outwardly propagating wave with a much shorter radial wavelength than the mode visualized in Fig. \ref{fig:Af1c02r18_composite}. PSDs analogous to those shown in Fig.~\ref{fig:Af1c02r18_mflp_psd} reveal a nearly identical frequency of $\omega\lesssim\max[\kappa]$, which implies a corotation radius of $r\sim11r_g$ (indicated by the white dashed line). Negligible amplitudes at this corotation radius indicate a damping of the outwardly propagating wave, which feeds back into the r-mode excitation mechanism (see Section \ref{sec:bgEx}).

In addition to axisymmetric and non-axisymmetric inertial waves, Figs. \ref{fig:Af1c02r18_composite}-\ref{fig:Af1c02r18_mflp_psd} indicate the presence of a third noteworthy dynamical feature: in  Fig. \ref{fig:Af1c02r18_composite}, secondary oscillations are faintly visible as vertical columns to the right of the trapped r-mode. These columns correspond to axisymmetric inertial-acoustic f-modes, and the space-time diagrams for f07c2, f10c2 and f10c1 illustrate their outward propagation (see the diagonal lines in Figs. \ref{fig:mflpr_spc} and \ref{fig:mflpr_spc1}). In Fig. \ref{fig:Af1c02r18_mflp_psd} (left), the f-modes correspond to a weak band in power at radii $r\gtrsim 9r_g$ with nearly double the trapped inertial wave's frequency. This is shown by the cyan line in Fig. \ref{fig:Af1c02r18_mflp_psd} (right), which gives ten times the power integrated over radii from $r=10r_g$ to the outer boundary. 

These vertically homogeneous f-modes cannot be attributed to `leakage' or `tunneling' of the trapped inertial waves out of their effective potential well, as they do not share the same vertical wavenumber. Moreover, they only begin to propagate after the growth of the trapped inertial waves, which suggests that they may be the result of an additional non-linear coupling. The most likely possibility is that this follows a self-interaction of the r-mode; the standing mode formed within the trapping region involves a superposition of oscillations with frequency $\omega\approx\max[\kappa]$, vertical wavenumber $k_z\sim\pm\pi/H$, and azimuthal wavenumber $m=0.$ Their non-linear interaction might then be expected to produce an f-mode with $\omega_f\approx2\max[\kappa]$, $k_f\approx\pi/H-\pi/H=0$ and $m_f=0.$ Examination of the space-time diagrams in Figs. \ref{fig:mflpr_spc} and \ref{fig:mflpr_spc1} shows that axisymmetric f-mode excitation accompanies and may aid in the saturation of r-mode growth. This non-linear interaction of modes with different physical character is interesting in light of observations of multiple HFQPOs with frequencies in near-integer ratios in some sources, although the inclusion of magnetic fields and density stratification will change the relevant mode couplings.

\subsection{Periodic vertical boundary conditions}\label{sec:pUBC}
Lastly, we mention f10c2p, a simulation run using the same setup as f10c2 but with purely periodic vertical boundary conditions imposed at $z=\pm H$ (rather than $u_z=0$). The lightest lines in Figs. \ref{fig:htrKE} (top) illustrate a similar evolution of vertical kinetic energy between f10c2p and f10c2. Simulation f10c2p also initially exhibits the growth of an oscillation in the expected trapping region, as illustrated by the space-time diagram showing azimuthally averaged radial mass flux in Fig. \ref{fig:mflpr_spc} (bottom).

\begin{figure*}
    \centering
    \includegraphics[width=\textwidth]{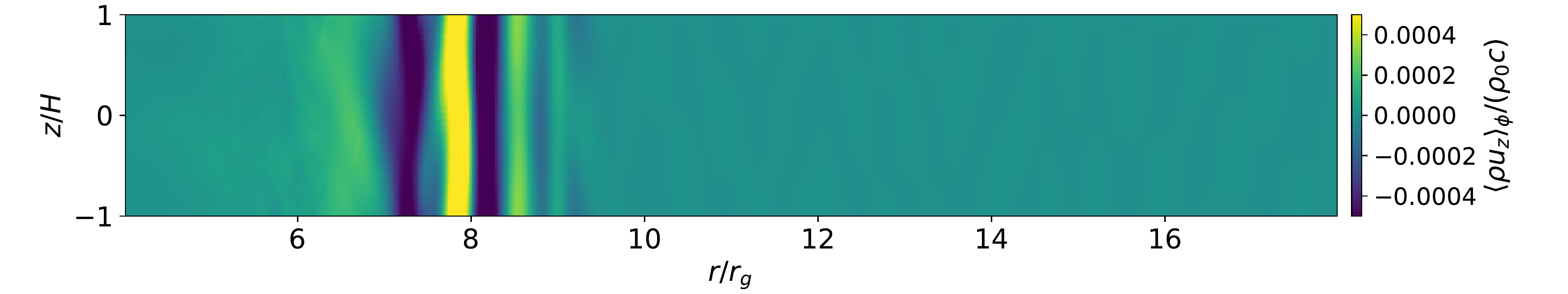}
    \caption{Meridional snapshot of azimuthally averaged vertical mass flux in the simulation f10c2p (taken at $t\omega_g\sim 3700$), which was run with periodic vertical boundary conditions. The colour-plot shows a steady `elevator' flow, a numerical artefact of the periodic vertical boundary conditions, which has halted the growth of the r-mode.}
    \label{fig:Af1c02r18_pUBC1}
\end{figure*}

However, the periodicity at later times ($t\gtrsim2000\omega_g^{-1}$) is altered, due to the formation of mean vertical flows as shown in Fig. \ref{fig:Af1c02r18_pUBC1}. These `elevator flows' are named as such because they appear in the azimuthally averaged flow as steady columns or `elevators' of constant $u_z$. Examination of snapshots suggests that they originate in the interaction between incident waves that travel upward and downward in the absence of vertical gravity. In the simulations with $u_z$ set to zero in the ghost cells, the incident waves self-organize into global modes like the one shown in Fig. \ref{fig:Af1c02r18_composite}, but with periodic BCs they simply shear out and form a vertically homogeneous pattern. The elevator flows may point to an attempt by the dynamics to create a large-scale circulation. Such circulation could be physically meaningful, but cannot be appropriately captured in a cylindrical model. While the vertically local framework used in this paper is sufficient to demonstrate r-mode excitation via non-linear coupling with disc eccentricity, a true characterization of the saturated state may require more expensive simulations fully incorporating vertical gravity and density stratification.

\section{Conclusions}\label{sec:conc}

In this paper, we have explored the excitation of trapped inertial waves, one explanation for the high-frequency, quasi-periodic oscillations that appear in the emission of black hole X-ray binary systems. Expanding upon previous three-wave coupling analyses of the problem, we have run global, hydrodynamic simulations of eccentric, relativistic discs. These simulations exhibit trapped r-mode excitation in the presence of sufficiently non-circular streamlines. The `very high' emission states in which HFQPOs are observed may provide a favorable environment for such non-circularity, since high accretion rates can aid the inward propagation of eccentricity from the outer disc, where it may be excited by the stellar companion \citep{fer09}. 

We employ a cylindrical, unstratified framework without vertical gravity, in which the saturation of inertial wave growth depends on the vertical boundary condition; global standing modes form if vertical velocity is set to zero in the ghost cells, but with periodic vertical boundary conditions the formation of elevator flows (mean vertical flows that are steady and vertically constant, but vary radially) precludes this outcome. The latter behaviour we consider an artifact of the numerical set-up.
The saturation of the trapped mode growth involves two processes: (i) the time-dependent reconfiguration of the trapping region (and mode oscillation frequency), and (ii) the secondary excitation of axisymmetric inertial-acoustic (density) waves.

In the presence of any dissipation, trapped inertial waves can transport angular momentum locally, and upon attaining sufficient amplitude can re-organise the background orbital flow and in particular the local epicyclic frequency; in some of our simulations this re-organisation takes the form of a quasi-steady zonal flow. By creating such zonal flows, trapped modes can re-shape the trapping region confining them. This process is dynamic and time-dependent, leading to an oscillation frequency that continually shifts, blurring the signal and perhaps partially accounting for HFQPOs' low quality factors.

The secondary inertial-acoustic waves result from a non-linear interaction between trapped inertial modes with vertical wavenumbers of equal amplitude and opposite sign. The appearance of distinct frequencies due to non-linear mode coupling is intriguing in light of observations of multiple HFQPOs in some sources. As a consequence of wave coupling rules, the density waves exhibited by our simulations are excited with a frequency roughly twice that of the trapped r-modes. This differs from the often mentioned ratio of 3:2 \citep[e.g.,][]{Mot14}, although the peaks in power for the f-modes vary over a broad enough range in frequency so as not to be inconsistent with this commensurability (see Fig. \ref{fig:Af1c02r18_mflp_psd}, right).

In any case, the analogous coupling of r-modes in a fully global framework including vertical gravity will likely differ in several respects, since in a density stratified model standing modes can no longer be considered as the superposition of Fourier components with positive and negative wavenumbers. The non-linear coupling of fully global trapped r-modes nearing saturation may then produce additional oscillations with different vertical structures and frequencies than the f-modes described in Section \ref{sec:3D}. Fully exploring the possibility that multiple HFQPOs might result from the non-linear interaction of discoseismic modes will  likely require more numerically expensive simulations that include vertical gravity. While sufficient for capturing the essentials of the r-mode excitation mechanism, the cylindrical model utilized in our simulations also excludes vertical resonances and a potential for r-mode coupling with warping disc deformations. The necessity of fully 3D simulations is further indicated by the growth of the artificial elevator flows when periodic vertical boundary conditions are used.

Several other aspects of the discoseismic model for HFQPOs remain to be fully explored. Most importantly, the extent to which r-modes are \emph{actively} damped by turbulent fluctuations must be quantified, alongside the capacity of the excitation mechanism to overcome such damping. The isothermal model considered here is also oversimplified; more realistic thermodynamics, with radiative physics, should be included to produce a better and more representative model. Lastly, it remains to be determined how dynamical oscillations in the disc might be communicated to the corona, and whether they can force its quasi-periodic, higher-energy emission. A theory outlining this communication is necessary to connect discoseismic models directly to observations, since HFQPOs appear imprinted on the high-energy, power-law (rather than the thermal) component of X-ray binaries' emission.

The simulations presented in this paper demonstrate that even very mildly eccentric relativistic discs support discoseismic mode excitation. The viability of trapped inertial waves as an explanation for HFQPOs likely depends most heavily on the robustness of their excitation in the presence of MRI turbulence. We explore this competition between damping and driving through global, MHD simulations of eccentric, relativistic discs in a companion paper. 

\section*{Acknowledgements}
The authors thank the reviewer for very useful comments and suggestions, which improved the quality of the paper. J. Dewberry thanks Adrian Barker, Roman Rafikov, Pascale Garaud, and Omer Blaes for helpful discussions. This work was funded by the Cambridge Commonwealth, European and International Trust, the Vassar College De Golier Trust, the Cambridge Philosophical Society and the Tsung-Dao Lee Institute.

\section*{Data availability}
The data underlying this article will be shared on reasonable request to the corresponding author.



\bibliographystyle{mnras}
\bibliography{trExHsim} 




\appendix

\section{Test simulations}\label{sec:tests}
In this Appendix we present axisymmetric ($r-z$) and purely 2D ($r-\phi$) simulations run to establish a baseline and test our version of RAMSES. These test simulations utilize quasi-rigid boundary conditions at the radial boundaries; while radial and vertical velocities are set to zero, the density in the ghost cells is matched to the value in the last cell in the active domain. Meanwhile, the azimuthal velocity is matched to the (extrapolated) equilibrium rotation defined by Equation \eqref{eq:1dEqm}. Note that in these test simulations we have also placed the inner boundary $r_0$ at the ISCO ($r_\text{ISCO}=6r_g$), rather than at the radius of $4r_g$ used in the runs described in Section \ref{sec:3D}.

\subsection{Linear r-mode evolution}\label{app:lintr}
To provide a point of comparison for our simulations following the non-linear excitation of r-modes, we have run simulations to verify their linear evolution. Table \ref{tab:lTRsim} summarizes axisymmetric simulations initialized with a hydrodynamic trapped inertial wave inserted by hand. The simulations are axisymmetric in that the fields depend only on $r$ and $z$. Aside from a change in the location of the inner boundary to $r_0=r_\text{ISCO}=6r_g$, we set up these runs to match those described in Section \ref{sec:3D}; we again choose a vertical domain $z\in[-H,H]$, use equivalent resolutions, and implement the same rigid vertical boundary conditions described in Section \ref{sec:BC}.

We generate linear oscillations to impose in the initial conditions using the method of calculating cylindrical, hydrodynamic modes described in \cite{dew18}. We initialize standing modes, with perturbations $\delta X(r,z)$ to a given variable $X$ calculated as $\delta X(r,z)=Re[\delta X(r,k_z)e^{\text{i}k_zz} + \delta X(r,-k_z)e^{-\text{i}k_zz}]/2$. Here the $\delta X(r,\pm k_z)$ are Fourier components calculated with a vertical wavenumber $k_z=\pm \pi/H$ chosen so that one wavelength of the oscillation spans the vertical domain. These standing modes are superimposed on top of the pseudo-Keplerian flow with a relative amplitude of $<10^{-4}$ compared to the background. 

\begin{table}
\centering
 \caption{Table listing (i) isothermal sound speed, (ii) resolution, (iv) frequency predicted by linear solver, v) frequency measured from the initial period of oscillation for axisymmetric simulations in $(r,z)$ that have been initialized with linear trapped inertial waves. Both simulations were run with domains $[r_0,r_1]\times[z_0,z_1]=[6r_g,18r_g]\times[-H,H].$}\label{tab:lTRsim}
\begin{tabular}{lccr} 
    \hline
    $c_s/c$  & 
    $N_r\times N_z$ & 
    $\omega_T/\omega_g$ & 
    $\omega/\omega_g$ \\
    \hline
    $0.02$  & $456\times 32$ & $0.03423$ & $0.03422$ \\
    $0.01$  & $900\times 32$ & $0.03445$ & $0.03461$ \\
  \hline
 \end{tabular}
\end{table}

\begin{figure}
    \centering
    \includegraphics[width=\columnwidth]{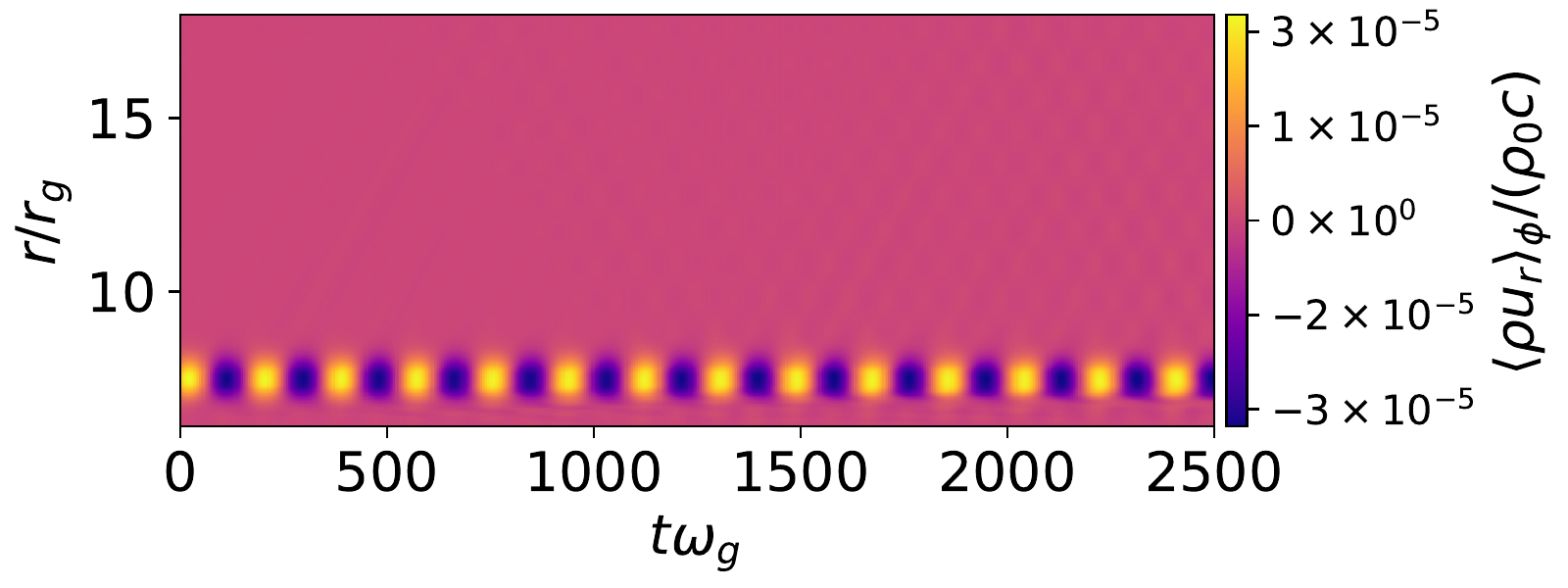}
    \caption{Space-time diagram showing mid-plane radial mass flux over time for the simulation initialized with a linear r-mode and $c_s=0.02c.$}
    \label{fig:ltr_uruzsp}
\end{figure}

The space-time diagram in Fig. \ref{fig:ltr_uruzsp} plots radial profiles of mid-plane radial mass flux over time for the simulation with $c_s=0.02c$. The plot may be compared with the space-time diagrams given in Fig. \ref{fig:mflpr_spc}, noting that these axisymmetric simulations do not include the inner, evacuated region present in our primary runs. Fig. \ref{fig:ltr_uruzsp} illustrates oscillatory behavior very similar to that exhibited by the 3D simulations described in Section \ref{sec:3D}, although in an isolated context the purely oscillatory r-modes remain better trapped, and maintain more coherent frequencies. These frequencies match the predictions of the linear calculations very closely (see Table \ref{tab:lTRsim}).

As mentioned in the text, the linear trapped inertial waves produce weak, vertically homogeneous `zonal flows' via a local redistribution of angular momentum. This process may be caused by the intrinsic momentum transport of the oscillation itself through its Reynolds stress. Ideal linear inertial modes possess perturbations in $u_r$ and $u_\phi$ that are exactly out of phase and thus no momentum flux is possible. But damping (in this case by numerical viscosity) will disturb the phase relation between the velocity perturbations, producing a non-zero flux and causing the growth of a zonal flow over time.

\begin{figure}
    \centering
    \includegraphics[width=\columnwidth]{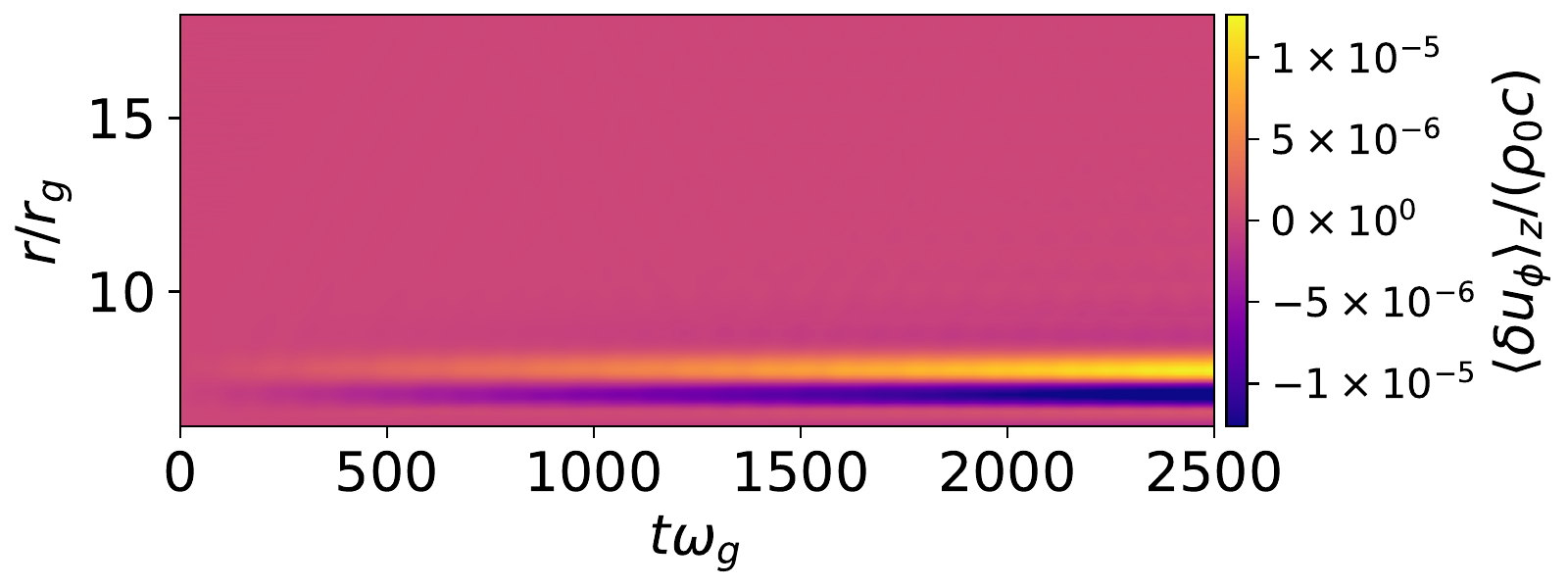}
    \caption{Spacetime diagram showing radial profiles of the vertically averaged fluctuations from equilibrium for azimuthal velocity in the linear eigenmode simulation with $c_s=0.02c$. This vertical average is calculated as $\int (u_\phi-r\Omega)\text{d} z/\int \text{d} z.$}
    \label{fig:ltr_dRdUpSp}
\end{figure}

Fig. \ref{fig:ltr_dRdUpSp} illustrates the growth of a zonal flow in the simulation with $c_s=0.02c$. The space-time diagram in the figure shows a vertical average of the fluctuation from the background equilibrium for the azimuthal velocity. Since the linear r-modes are periodic in $z$, this vertical average removes the oscillation and leaves only the growing zonal flow. The amplitudes of the zonal flows produced by the linear r-modes considered in this section are inconsequential. Given sufficient amplitude, however, zonal flows can play a more significant role in altering flow dynamics \citep[see, e.g.,][]{wei18}. The most relevant effect to the simulations considered in Section \ref{sec:3D} is a modification of the profile for $\kappa$.

Contained within their trapping cavities, linear r-modes are not significantly affected by radial boundary conditions. While some wave energy is lost through leakage and axisymmetric f-mode excitation, the decay in r-mode amplitudes is negligible over hundreds of inner orbital periods. The modes do show some interaction with an inner boundary placed at the ISCO, however. While the oscillation frequencies themselves remain unaffected, this interaction provides partial motivation for our choice to allow material to flow through the ISCO in our main simulations.

\subsection{Spiral density wave excitation}\label{sec:corot}
To test the performance of our code in the non-linear regime and over the full azimuthal range $\phi\in[0,2\pi)$, we have simulated the growth of spiral density waves in relativistic discs via the corotation instability 
\citep{lai09,FL11}. The numerical results of these simulations are closely comparable with those of \cite{FL13}.

\begin{figure}
    \centering
    \includegraphics[width=\columnwidth]{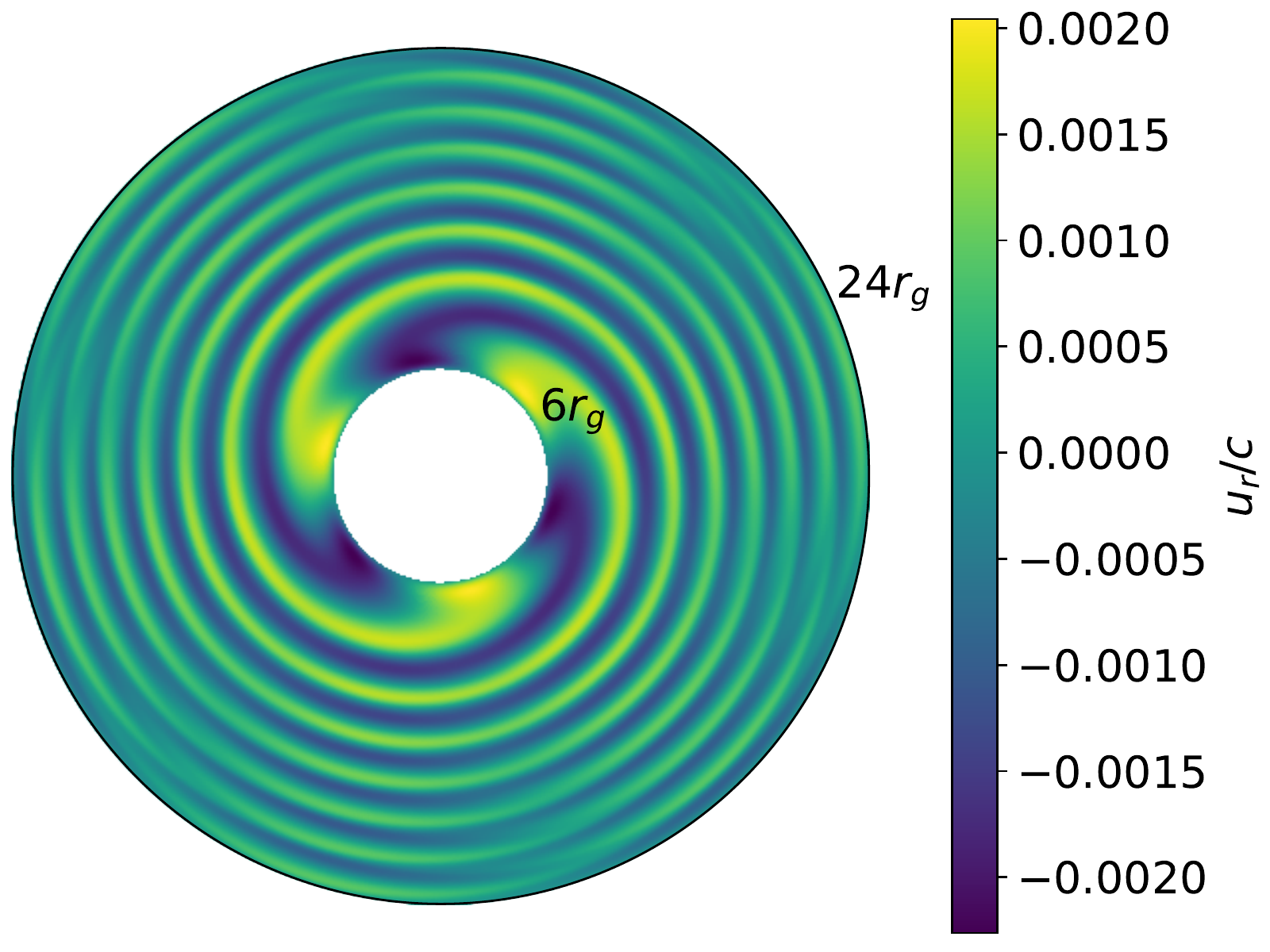}
    \caption{Snapshot of radial velocity taken at $t=25T_\text{orb}$ in simulation m3, illustrating the growth of an $m=3$ f-mode.}
    \label{fig:corot} 
\end{figure}

Table \ref{tab:CRsim} summarizes the relevant parameters used in these test simulations, which have been chosen to mirror those of \cite{FL13}. To allow continued outward wave propagation, a non-reflective, wave damping outer boundary condition \citep{deva06} at $r_1$ supplements the quasi-rigid inner boundary condition used in the previous section. Upon initialization, we seed the corotational instability with density perturbations $\delta \Sigma\lesssim 10^{-4}\Sigma_0$ that are random in radius and have either $m=2$, $m=3$ or random structure in azimuth. The growth rates listed in Table \ref{tab:CRsim} have been calculated from time series tracking the maximum amplitude in radial velocity. We find growth rates comparable to those found by \cite{FL13}, and similarly observe saturation of the instability when the maximum radial velocity nears sonic values. Fig. \ref{fig:corot} shows a snapshot of radial velocity from the simulation initialized with $m=3$ perturbations, illustrating the growth of an $m=3$ f-mode. 
\begin{table}
\centering
\caption{Table of parameters describing simulations run to follow the growth of inertial-acoustic modes excited by the corotation instability. All three simulations have been run with a density profile $\Sigma\propto r^{-1},$ a sound speed $c_s\approx 0.06c,$ a simulation domain $[6r_g,24r_g]\times[0,2\pi),$ and a resolution of $N_r\times N_\phi=1024\times2048$. The azimuthal wavenumber $m_p$ describes the azimuthal structure of initial density perturbations that are random in radius ($m_p=$None corresponds to a perturbation random in both $r$ and $\phi$). Meanwhile, $s$ and $s_\text{FL13}$ are the growth rates of inertial-acoustic (spiral density) waves measured in our simulations and by \protect\cite{FL13}.}\label{tab:CRsim}

\begin{tabular}{lccr} 
    \hline
    Name &
    $m_p$  &   
    $s/\omega_g$ &
    $s_\text{FL13}/\omega_g$\\
    \hline
    m0 & None  & $6.1\times10^{-3}$ & $7.5\times10^{-3}$ \\
    m2 & 2     & $5.9\times10^{-3}$ & $6.5\times10^{-3}$ \\
    m3 & 3     & $5.9\times10^{-3}$ & $7.5\times10^{-3}$ \\
  \hline
 \end{tabular}
\end{table}

The corotational instability significantly disrupts linear eccentric modes \citep[calculated, e.g., as described by ][]{fer08} initialized in 2D simulations with a rigid boundary placed at the ISCO, even with the use of a buffer zone that relaxes the flow toward the disc eccentricity on a dynamical timescale. However, in agreement with \cite{mir15} we find that the corotational instability only operates in the presence of a reflective inner boundary: growth rates are greatly reduced by a zero-gradient inner boundary condition, and the non-axisymmetric f-modes disappear entirely when the inner boundary $r_0$ is placed within the ISCO.

\section{Resolution study}\label{app:conv}
To check that the resolutions used in Section \ref{sec:3D} are sufficient to capture r-mode excitation, we have run simulations with the same parameters as f10c2, but with the number of grid points both halved and doubled in each (and every) direction. Table \ref{tab:restudy} lists the labels, resolutions and estimated r-mode growth rates associated with these simulations. Notably, simulations f1c2mll and f1c2hmm, which have more azimuthally elongated grid cells, do show depressed r-mode growth rates. For aspect ratios $r\text{d}\phi/\text{d}r\lesssim3$ at the ISCO, however, these growth rates return to the values measured for f10c2. 

\begin{figure}
    \centering
    \includegraphics[width=\columnwidth]{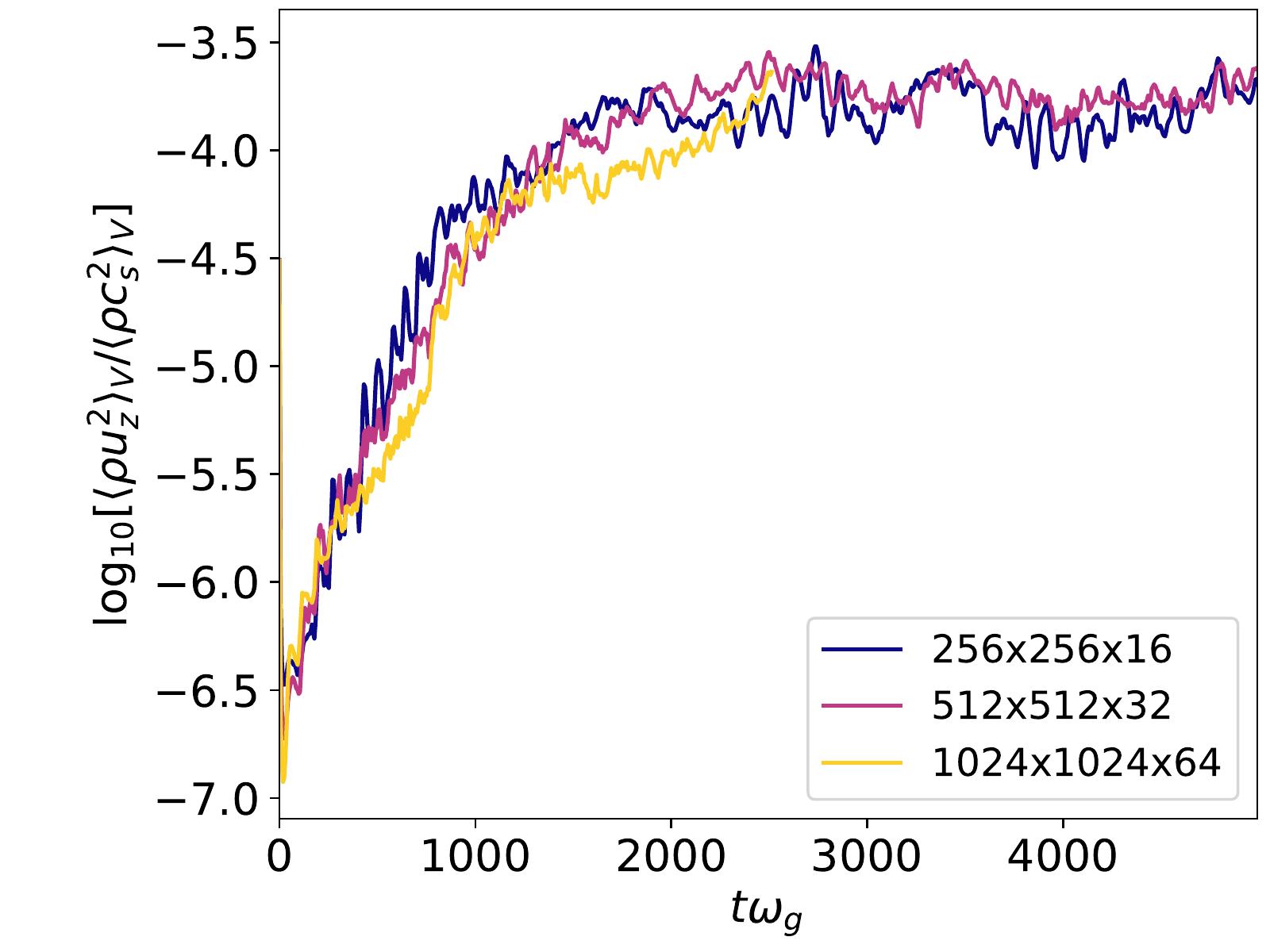}
    \includegraphics[width=\columnwidth]{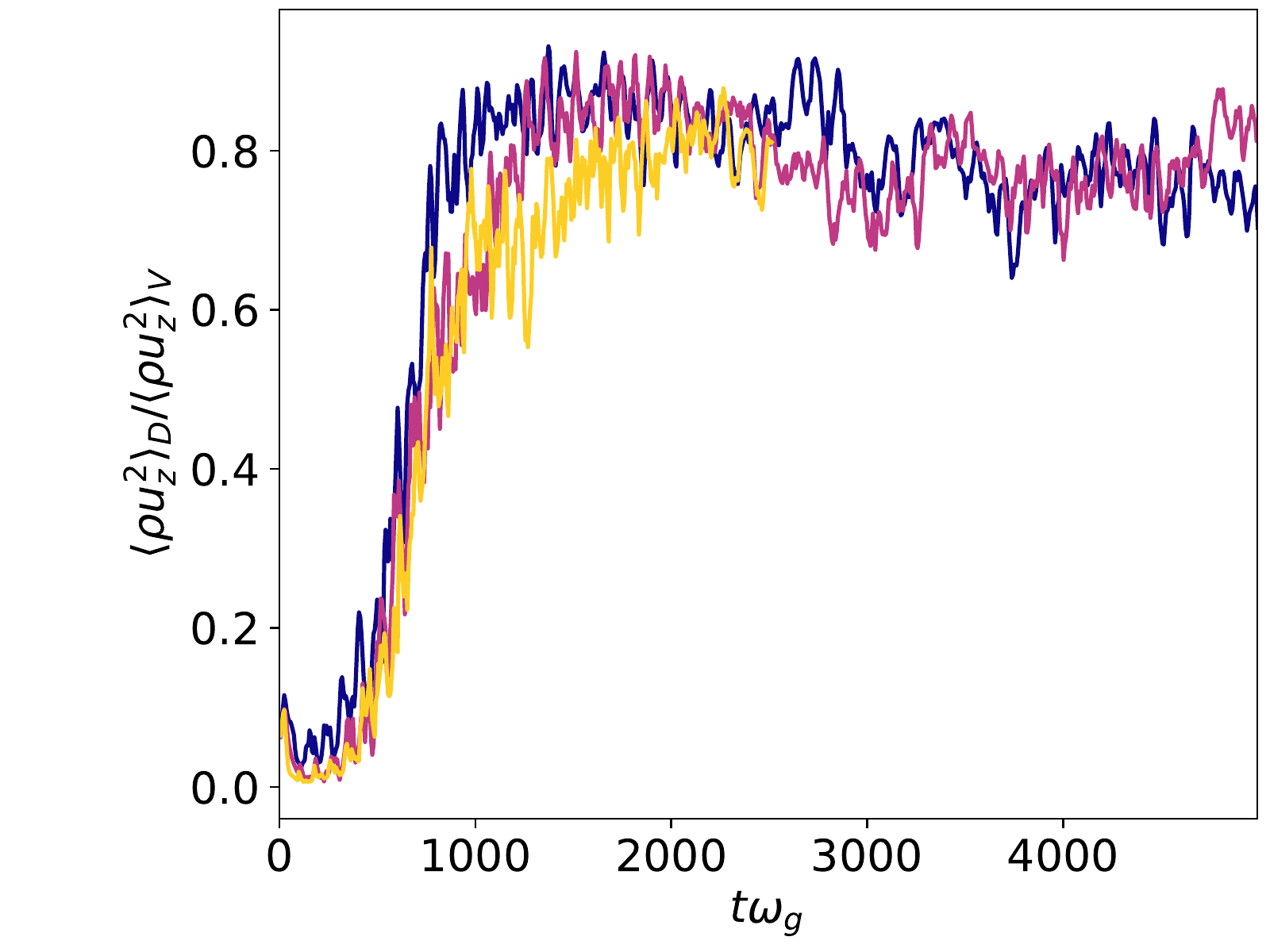}
    \caption{Top: volume averaged vertical kinetic energy density for three of the simulations listed in Table \ref{tab:restudy}. Bottom: fraction of vertical kinetic energy contained within the annulus defined by $r\in[7r_g,9r_g].$} \label{fig:resKE}
\end{figure}

Like Fig. \ref{fig:htrKE}, Fig. \ref{fig:resKE} plots the total average of vertical kinetic energy (top) and the fraction contained between $7r_g$ and $9r_g$ (bottom) for simulations f1c2lll, f10c2 and f1c2hhh (the three simulations in which resolutions are altered in all three dimensions simultaneously). The plots in Fig. \ref{fig:resKE} indicate that increasing resolution has a minimal impact on r-mode excitation.

\begin{table}
\centering
\caption{Table listing the (i) label, (ii) resolution, and (iii) estimated r-mode growth rate for simulations run with the same parameters as f10c2 ($A_f=0.1,$ $e_\text{tr}\sim0.013,$ $c_s=0.02c$, simulation domain $[4r_g,18r_g]\times[0,2\pi)\times[-H,H]$) but different grid resolutions. All simulations were run for $5000\omega_g^{-1}$ except for f10c2hhh, which was run for $2500\omega_g^{-1}$.}\label{tab:restudy}
\begin{tabular}{lcr} 
    \hline
    Label    &
    $N_r\times N_{\phi}\times N_z$ &
    $s/\omega_g$ \\
    \hline
    f1c2lll & $256\times 256\times 16$  & $5.7\times 10^{-3}$ \\
    f1c2llm & $256\times 256\times 32$  & $6.2\times 10^{-3}$ \\
    f1c2lml & $256\times 512\times 16$  & $5.8\times 10^{-3}$ \\
    f1c2mll & $512\times 256\times 16$  & $4.8\times 10^{-3}$ \\
    f10c2   & $512\times 512\times 32$  & $5.8\times 10^{-3}$ \\
    f1c2mmh & $512\times 512\times 64$  & $5.6\times 10^{-4}$ \\
    f1c2mhm & $512\times 1024\times32$  & $5.3\times 10^{-3}$ \\
    f1c2hmm & $1024\times512\times 32$  & $4.7\times 10^{-3}$ \\
    f1c2hhh & $1024\times1024\times 64$ & $5.5\times 10^{-3}$ \\
  \hline
 \end{tabular}
\end{table}


\bsp	
\label{lastpage}
\end{document}